\documentclass[aps,twocolumn,preprintnumbers,amsmath,amssymb, superscriptaddress, showkeys,nofootinbib]{revtex4-1}
\pdfoutput=1
\usepackage{amsmath,amsfonts,amstext,txfonts,amssymb,rotating,
dcolumn,graphics,array,multirow,sidecap,mathrsfs,gensymb,graphicx,units,hyperref}
\newcommand{\comillas}[1]{\textquotedblleft #1\textquotedblright}

\DeclareMathOperator{\Tr}{Tr}
\DeclareMathOperator{\sinc}{sinc}

\begin{document}
\title{Temporal correlations of sunlight may assist photoprotection in bacterial Photosynthesis}

\author{Adriana M. De Mendoza}
\affiliation{Physics Department, Universidad de Los Andes,  A.A. 4976  Bogot\'a, Colombia}
\affiliation{OncoRay – National Center for Radiation Research in Oncology, Faculty of Medicine and University Hospital C. G. Carus, TU Dresden, HZDR, Dresden, Germany}

\author{Felipe Caycedo-Soler}
\affiliation{Institut f\"ur Theoretische Physik and IQST, Universit\"at Ulm, Albert-Einstein-Allee 11, 89081 Ulm, Germany}

\author{Susana F. Huelga}
\affiliation{Institut f\"ur Theoretische Physik and IQST, Universit\"at Ulm, Albert-Einstein-Allee 11, 89081 Ulm, Germany}

\author{Martin B. Plenio}
\affiliation{Institut f\"ur Theoretische Physik and IQST, Universit\"at Ulm, Albert-Einstein-Allee 11, 89081 Ulm, Germany}
\begin{abstract}{
Photosynthetic systems utilize adaptability to respond efficiently to fluctuations in their light environment.  As a result, large photosynthetic yields can be achieved in conditions of low light intensity, while photoprotection mechanisms are activated in conditions of elevated light intensity. In sharp contrast with these observations, current theoretical models predict bacterial cell death for physiologically high light intensities. To resolve this discrepancy, we consider a unified framework to describe three stages of photosynthesis in natural conditions, namely light absorption, exciton transfer and charge separation dynamics, to investigate the relationship between the statistical features of thermal light and the Quinol production	 in bacterial photosynthesis. This approach allows us to identify a mechanism of photoprotection that relies on charge recombination facilitated by the photon bunching statistics characteristic of thermal sunlight. Our results suggest that the flexible design underpinning natural photosynthesis may therefore rely on exploiting the temporal correlations of thermal light, manifested in photo-bunching patterns, which are preserved for excitations reaching the reaction center.
}
\end{abstract}

\maketitle


Natural photosynthesis thrives under constantly changing environmental conditions thanks to active regulatory mechanisms.  These mechanisms must balance the interplay between  
thermal light absorption, transport of the resulting electronic excitations (excitons) to a reaction center (RC), charge separation and metabolic output. However, the sheer complexity of these four individual processes makes their combined description challenging and fundamental questions remain open. In particular, photosynthetic bacteria can survive in light intensities that exceed predictions from current theoretical models\cite{Scheuring,Geyer,Cogdell1,Kim,Fleming1,Kosumi,Scully-blockade}. To resolve this discrepancy, we propose a novel approach to describe the coupled dynamics resulting from those fundamental processes 
in real photosynthetic vesicles, as depicted in Fig.\ref{photos}.

\textbf{Illumination}. The  coherence length of sunlight at the earth's surface is approximately 50 $\mu$m \cite{Bures71,Mashaal2}. This value exceeds the size of any unicellular photosynthetic organism, e.g. a purple bacterium, as illustrated in Fig.\ref{photos}(a). Hence, the optically active molecules of photosynthesis, which arrange in light harvesting complexes (LH1 and LH2 in purple bacteria) within photosynthetic vesicles, will absorb thermal photons with significant temporal correlations leading to interspersed bursts of photons, as noted decades ago \cite{Kano,Glauber_1963PR,Mandel_1964PPS}.

\textbf{Exciton transfer}. After photon absorptions, the resulting excitons diffuse across the network of LHs until reaching an available RC within a LH1 
 (see Fig.\ref{photos}(b). The transfer rates $t_{ij}$ between complexes $i\rightarrow j$ are adequately described via generalized F\"orster theory \cite{Silbey_JCP2003} and accurately tested by linear and multidimensional spectroscopic techniques \cite{Bergstrom_1989FEBS,Timpmann,Vgrondelle,Visscher_1989PhotRes,Engel_PNAS2012,Ogilvie_NChem2014,Zigmantas_2018NatChem,vanGrondelle2}.

\textbf{Charge separation dynamics}. Once an exciton arrives at a RC it triggers an electron transport chain, whose outcome is the production of quinol molecules ($Q_BH_2$). This transport goes through intermediate metastable states of charge separation with relevant recombination times ($t_{crit}$ in Fig.\ref{photos}(c)) \cite{Graige,Diner, Milano}, which in combination with the bursted structure of the incoming excitons may effectively reduce the conversion efficiency of the photosynthetic machinery. These processes are described in detail in the next section.

\textbf{Metabolism}. Quinol reduction facilitates the formation of an electrochemical gradient that drives the rotation of a transmembrane macromolecule ATP-synthase (ATPase) to synthesize Adenosine Triphosphate (ATP) (Fig.\ref{photos}(d)). The single ATPase present per vesicle may synthesise at most 100 molecules (about 200 excitations/s) of ATP for bacterial photosynthesis \cite{Geyer}. Based on this analysis, typical membranes with $\simeq$400-600 LHs \cite{Scheuring,vanGrondelle_PNAS2004} and 90\% transport efficiency \cite{Caycedo_Soler_2010NJP,Sener}, will accumulate hydrogen, increasing dangerously the cytoplasmic acidity for intensities as low as $\langle I \rangle $=0.1-0.2 Watt/m$^2$. This estimate is in stark contrast with the proven survival of purple bacteria for light intensities as high as 100 Watt/m$^2$ \cite{Scheuring,Sener}. In principle, charge recombination could provide the means to avoid the formation of excess quinol, but it has not been considered in previous attempts to provide a global description of the
primary steps in bacterial photosynthesis \cite{Caycedo_Soler_2010NJP,Caycedo_2010PRL,ADM1,Sener,review}.

In this article, we show that a detailed but comprehensive model of the elementary processes in the photosynthesis of purple bacteria unveils a dynamical interplay between the bunched statistics of thermal light, the process of charge recombination and quinol production. In particular, the long time-intervals between photon bursts resulting from thermal light absorption provide an essential element to determine the photoprotective role of metastable states in RCs.

\section{Results}

\subsection{Thermal light illumination and excitons statistics}

The photon-statistics of thermal light absorption shows typical intensity fluctuations that lead to patterns of bunched photo-detections separated by waiting times (where no detection happens), when the detection takes place within the  

\newpage
\onecolumngrid

\begin{figure}
	\includegraphics[width=0.9\columnwidth]{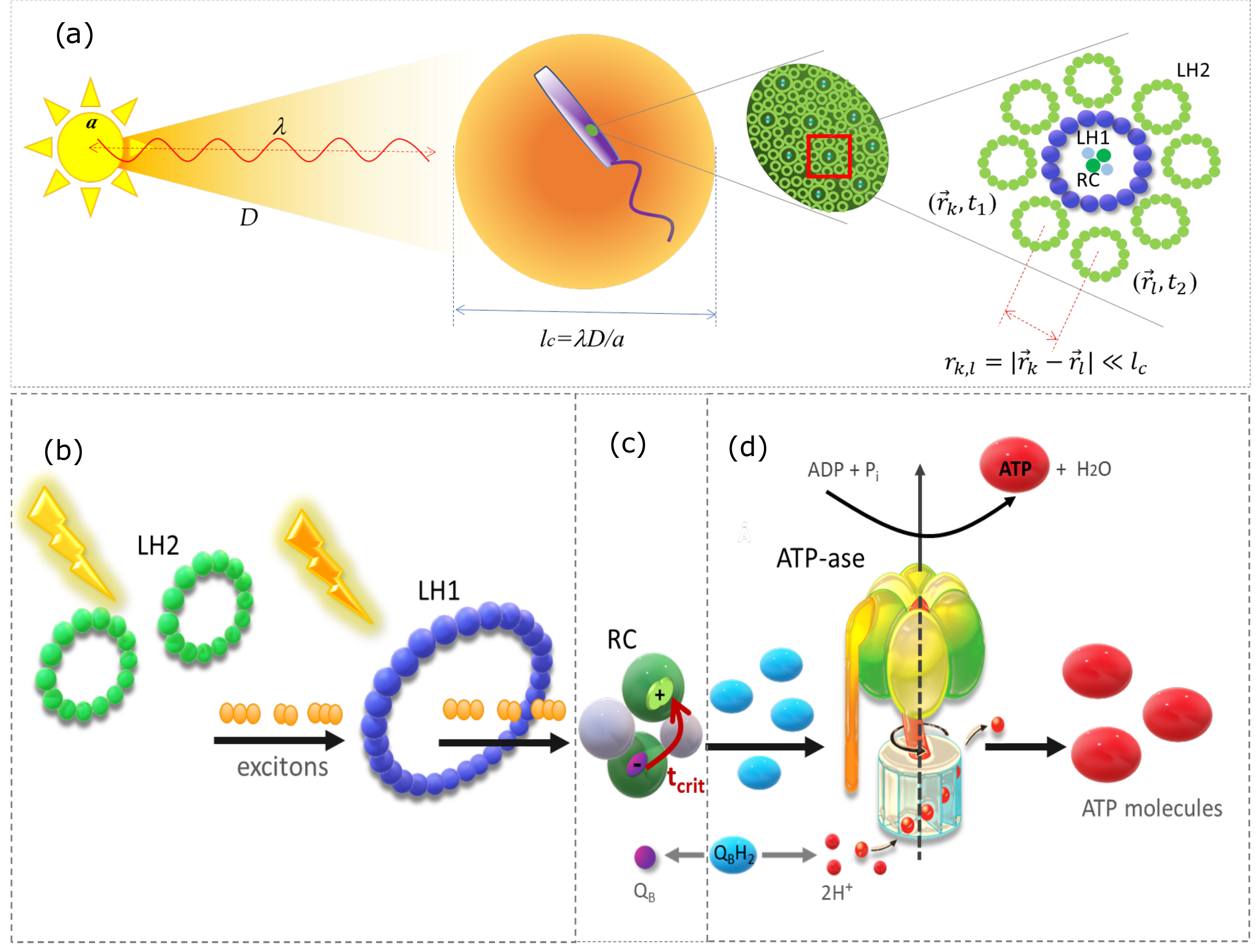}
				\caption{\textbf{(a) Relevant length scales in typical bacterial photosynthetic apparatus.} On the left side, we illustrate the parameters describing the incoherent light source provided by sunlight. Here $a$ denotes the source diameter, $D$ the distance to the light reception system and $\lambda$ the mean wavelength of the incident light. The large orange circle with diameter $l_c\approx 50 \mu$m, \cite{Mashaal2} represents the coherence area of sunlight, encircling a complete photosynthetic bacterium of length $\approx 40\mu$m \cite{Jungas}. It is within this coherence region defined by $l_c$ where thermal temporal correlations will become manifest. On the right side of panel (a) we depict the photosynthetic apparatus of the bacterium, embedded in characteristic vesicles. They contain the light harvesting complexes LH2 and LH1 and the encircled reaction centres (RCs), with typical inter-complex distance $r_{k,l}\ll l_c$.  \textbf{(b)-(d) Flow diagram illustrating the elementary processes in bacterial photosynthesis.} Photons (depicted as orange bolts) are detected by the light harvesting complexes (LH1 and LH2) and converted into photo-excitations totally or partially delocalized along individual harvesting complexes (excitons), which travel through the network of LHs until reaching an open RC as depicted in (b). Once in the RC, the exciton induces charge separation and a chain of electron transport steps that end up in quinol ($Q_BH_2$) production. Charge can be recombined, as represented by the red arrow, with recombination lifetime $t_{crit}$ in (c). In (d) quinol oxidizes and the liberated hydrogen (together with two additional hydrogens pumped from the cytoplasm) produce the electrochemical gradient required to make the ATPase macromolecule rotate and produce ATP. The ATPase molecule catalyzes the reaction between adenosine diphosphate (ADP) and inorganic phosphate (P$_i$).}
		\label{photos}
\end{figure}

\begin{figure}[t]
		\centering
			\includegraphics[width=.9\columnwidth]{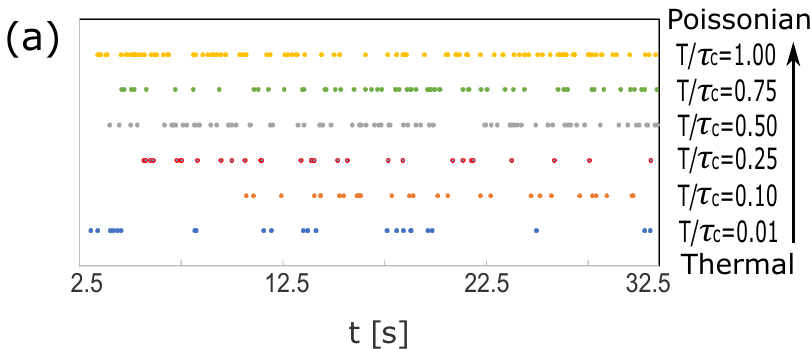}
			\includegraphics[width=1\columnwidth]{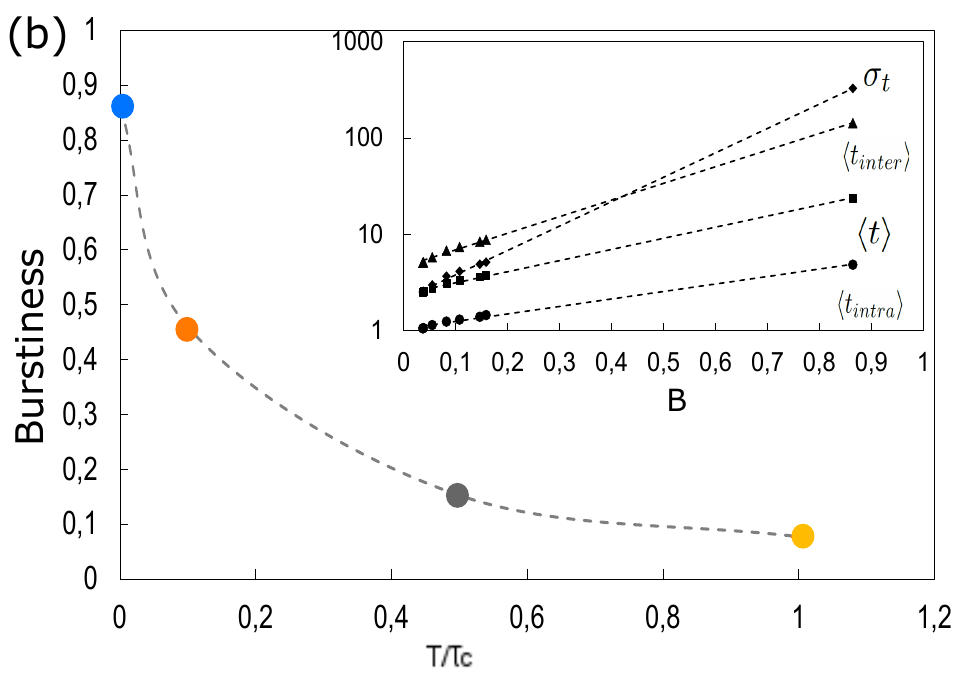}
		\caption{\textbf{Thermal versus Poissonian light.} (a) Typical time traces of absorption events for different values of the ratio $T/\tau_c$.	(b) Burstiness $B$ of the reception signal as a function of the temporal correlations ($T/\tau_c$). In the inset, we show the characterisation of low-B signals using as figure of merit the quantities $\langle t \rangle$, $\sigma_t$, $\langle t_{intra}\rangle$, and $\langle t_{inter}\rangle$ (descriptions in the text). All quantities are displayed in milliseconds on a semi-logarithmic scale. We considered$10^{6}$ light absorption events.}
		\label{incoming}
\end{figure}

\twocolumngrid

\noindent coherence length and coherence time of the light \cite{Rosseau_1975,Rosseau_1977,Kimble}. This pattern -- also called \textit{burstiness} of the detected photon-traces -- arises as a consequence of the light-matter interactions and hence depends on the joint properties of the light field (frequency spectrum, intensity and correlations) and the detection system provided by the LH1-2 complexes (absorption efficiency, absorption bandwidth, spatial configuration and detection time window). We have implemented a statistical model of thermal light detection where the probability distribution of absorption events is characterized by its factorial moments generating function \citep{ADM-IOP}. This framework, whose main features are revised in the Methods, allows for the inclusion of light correlations in a natural manner,  showing that the degree of burstiness can be tuned by the ratio between the detection time window and the coherence time of the detected light $T/\tau_c$. This parameter allows the continuous statistical tuning of the absorbed excitations from a thermal ($T/\tau_c\ll 1$) field with maximal burstiness, to a Poissonian field ($T/\tau_c\gtrsim 1$) without burstiness, as shown in Fig.\ref{incoming}(a). 

The statistics of bursts in a sequence of absorption events can be characterized by the so-called {\it burstiness} parameter, $B=(\sigma_t-\langle t\rangle)/(\sigma_t+\langle t\rangle)$. This quantity  vanishes in a fully Poissonian process where the mean inter-event time $\langle t\rangle$ equals the standard deviation of waiting times $\sigma_t$, but has a positive value whenever the times series exhibit bursts of events \cite{Barabasi}. Unsurprisingly, in Fig.\ref{incoming}(b)  we show that the ratio $T/\tau_c$ and $B$ are related monotonically. This implies that either of them can be used to quantify the bunching of the light absorption, and supports our use of $B$ in what follows as a measure of the burstiness in traces of events, in particular, on the arrival of excitations to the RCs.

To characterise the statistical properties of the waiting times $t$ between consecutive absorption events beyond the average $\langle t \rangle$ and variance $\sigma_t$, we consider the average waiting times between and within bursts, namely $\langle t_{inter}\rangle$ and  $\langle t_{intra}\rangle$, respectively. For the simulated events, all four statistical quantities increase as B increases and approaches 1 for a thermal distribution (see inset in Fig.\ref{incoming}(b)), while preserving the average intensity of the source (which is obtained by energy conservation and not by the reciprocal of $\langle t\rangle$ for thermal fields  \cite{Rockower}).
It is worth stressing from the inset in Figure \ref{incoming}(b) that $\sigma_t$ grows faster than $\langle t \rangle$. This reflects the faster growth of $\langle t_{inter}\rangle$ compared with $\langle t_{intra}\rangle$ for larger burstiness $B$. 
It can also be observed in  the same figure the quick convergence among  $\langle t \rangle$ and $\sigma_t$ for  $T/\tau_c\gtrapprox1$, as expected from the trend towards independent random events following a Poisson distribution.\\

Chan and collaborators recently showed that for the conditions of sunlight absorption, namely ultraweak chromophore–light coupling and intensity, the key parameter determining the quantum dynamics of absorption in photosynthetic systems is the ratio between the chromophores absorption spectral bandwidth $\Delta \omega$, and their concomitant emission rate $\Gamma$ \cite{Whaley}, which characterizes the recovery time of the pigments.

Since the coherence time of absorption events is approximated as the inverse of the spectral bandwidth of the absorption ($\tau_c \approx 2\pi/\Delta \omega$), and the recovery time of the chromophores sets the maximum of the detection time ($T_{max} \approx 1/\Gamma$); the ratio is analogous to the maximum of $T/\tau_c$, i.e., $\Delta \omega/\Gamma \approx 2\pi T_{max}/\tau_c$  (see Appendic A1 for further discussion). Regarding the recovery time, experiments have shown that excited state absorption occurs in LH complexes with about 90\% of the intensity of ground state absorption after $\sim$0.1 ps \cite{Trinkunas_PRL}. Therefore, at the chromophores absorption spectra ($\lambda_{LH1}=875\pm20$nm and $\lambda_{LH1}=850\pm20$nm) the coherence time ($\tau_c \sim 0.1$ps) is comparable to their recovery time ($1/\Gamma$), reflecting the similar energetic scales of exciton-exciton, exciton-phonon, phonon-phonon couplings and energetic disorder of this system \cite{Whaley}. In practice, due to the rather low physiological light intensity, doubly excited states will seldom arise in LHs and hence, the photosynthetic absorption set --composed of hundreds of LHs-- should be capable to discriminate individual photons within the coherence time. These results indicate that thermal light detection in photosynthetic systems might operate in a regime where the detection time is smaller or comparable to the coherence time, which makes it at least plausible that the temporal correlations present in thermal light may affect absorption statistics.

\subsection{Coupling thermal light absorption with exciton dynamics and charge separation on bacterial vesicles}

\begin{figure}
		\centering
		\includegraphics[width=1.0\columnwidth]{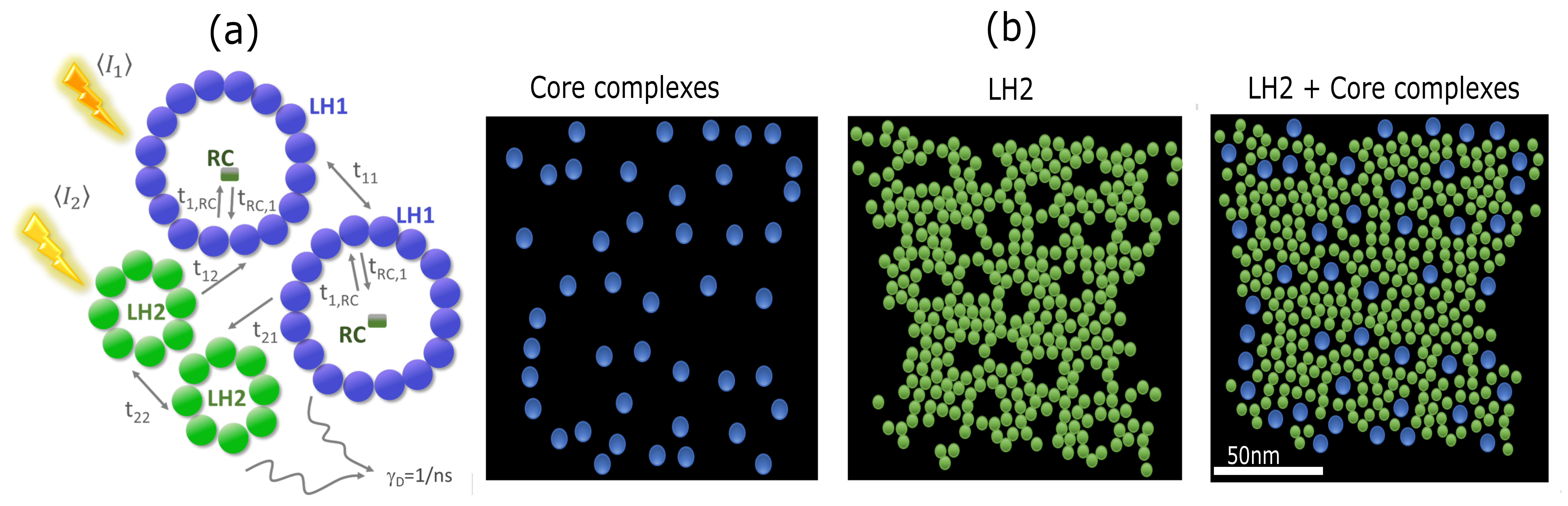}
		\caption{\textbf{(a) Description of the inter complexes transport dynamics.} After a photon absorption takes place, the exciton travels through the network of light harvesting complexes until arriving a RC. Here $t_{1,1}$ is the transfer rate between LH1s, $t_{2,2}$ labels the rate between LH2s, $t_{1,2}$($t_{2,1}$) denotes the transfer rate from an LH1 to an LH2 complex (viceversa). The transfer rate between an LH1 and its encircled RC is $t_{1,RC}$ and the respective reverse rate is $t_{RC,1}$. Excitonic transfer between LHs has an associated dissipation rate $\gamma_D$. (b) \textbf{Spatial arrangement of LH complexes in purple bacteria.} Configuration of 44 core complexes (LH1 + RC) (blue circles), and 356 LH2 (green cricles), for the intensity $\langle I\rangle=10^{-3}$photons/(LH$\cdot$ms) used in simulations.}
		\label{Trates1}
   \end{figure}

After photon absorption, the exciton dynamics across the membrane vesicle can be modeled based on the transfer rates between nearest neighbour complexes.
Figure \ref{Trates1} summarizes the excitation kinetics and spatial arrangements of the biomolecular complexes LH1, LH2 and RC in purple bacteria. In typical vesicles, the LH2 antenna outnumbers the LH1 complexes, which form an ellipse that  encircle the RC protein. Excitation transfer between RCs, LH1s and LH2s occurs from induced dipole transfer on a picosecond time-scale \cite{review}, while vibrational dephasing destroys coherences in hundreds of femtoseconds at most \cite{Engel_PNAS2012,Ogilvie_NChem2014,Zigmantas_2018NatChem,vanGrondelle2}, resulting in incoherent energy transfer between the RC$\leftrightarrows$LH1$\leftrightarrows$LH2 sustained, however, by delocalized excitons over single complexes \cite{Silbey_JCP2003,Bergstrom_1989FEBS,Timpmann,Vgrondelle,Visscher_1989PhotRes,Engel_PNAS2012,Ogilvie_NChem2014, Zigmantas_2018NatChem,vanGrondelle2,ADM-IOP,Hess,Mattioni}. Transfer rate measures from pump-probe experiments are available for  LH2$\leftrightarrow$ LH1 and  LH1$\leftrightarrow$ RC transfer steps \cite{Bergstrom_1989FEBS,Timpmann,Vgrondelle,Visscher_1989PhotRes,review}, whereas transfers  LH2$\leftrightarrow$LH2 and  LH1$\leftrightarrow$LH1, i.e. between the same type of complexes, have been estimated via generalized F\"orster rates \cite{ritz,review}.  Fluorescence, inter-system crossing, internal conversion and further dissipation mechanisms, have been included with a single conservative lifetime $1/\gamma_D$ of 1 ns \cite{ritz}. All transfer rates are depicted in Fig.3(a)
and their values are given in Table I.

\newpage
\onecolumngrid

\twocolumngrid

The formation of a \comillas{special pair} ($P$) in the RC initiates the first electron transfer step $P^*\rightarrow P^+$, followed by several reactions ending up with the reduction of the quinone $Q_A^-$ or auxiliary quinone $Q_B^-$. After $P$ becomes neutral following a rather fast cycle of $~1\mu$s, (See Appendix B for details on other photo-chemical transfer times scales involved) \cite{Graige,Osvath}, it can initiate a second electron transfer that results in the reduction of both quinones $Q_B^-Q_B^-$, which triggers the uptake of two intracytoplasmic protons to produce quinol $Q_BH_2$. The cycle starts over when $P$ is neutral again, and quinol is exchanged by $Q_B$ via affinity reactions \cite{Graige} (see Appendix B for a more detailed explanation). Previous work showed that the interplay between exciton dynamics and the cycling of quinol and the auxiliary quinone affects the quinol output in vesicles\cite{Caycedo_2010PRL,Caycedo_Soler_2010NJP}, which were observed to adapt to low and high light intensities \cite{Scheuring}. However, the actual output of quinol production in such a model is larger than the turnover of the ATPase molecular rotor discussed in the introduction, and would lead to bacterial death. The subsequent consideration of burst statistics in thermal light did not result in a notable reduction of the quinol output \cite{ADM1}, which is the achievement of the present work. Using the photon absorption time-traces exemplified in Fig.\ref{incoming}, we initiate the excitonic dynamics with the rates ($\gamma_D$, $t_{i,j}$ for $i$ or $j=$LH1,LH2,RC) depicted in Fig.\ref{Trates1}(a), over a typical photosynthetic vesicle  composed of a few hundreds of LHs, illustrated in Fig.\ref{Trates1}(b). In our simulation we include the charge dynamics occurring within the RCs, therefore, excitations reaching RCs initiate charge separation $P^*\rightarrow P^+$ and, crucially, going beyond previous considerations made in references  \cite{Caycedo_2010PRL,Caycedo_Soler_2010NJP,ADM1}, we also include the recombination of the intermediate metastable state $P^+QB^-$. This recombination takes place in about 100 ms to 1 s, a time-scale commensurable with photon absorption waiting times in physiological conditions. 

\begin{figure}
		\centering
		\includegraphics[width=1.\columnwidth]{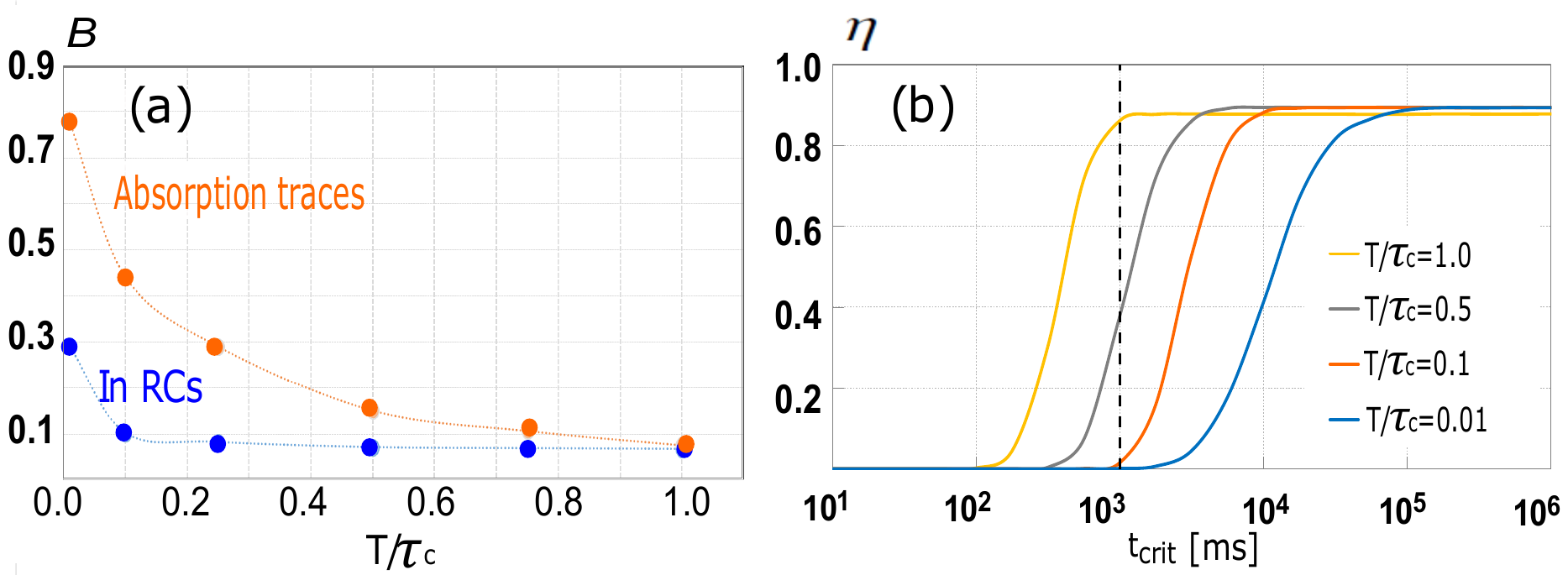}
			\includegraphics[width=1.\columnwidth]{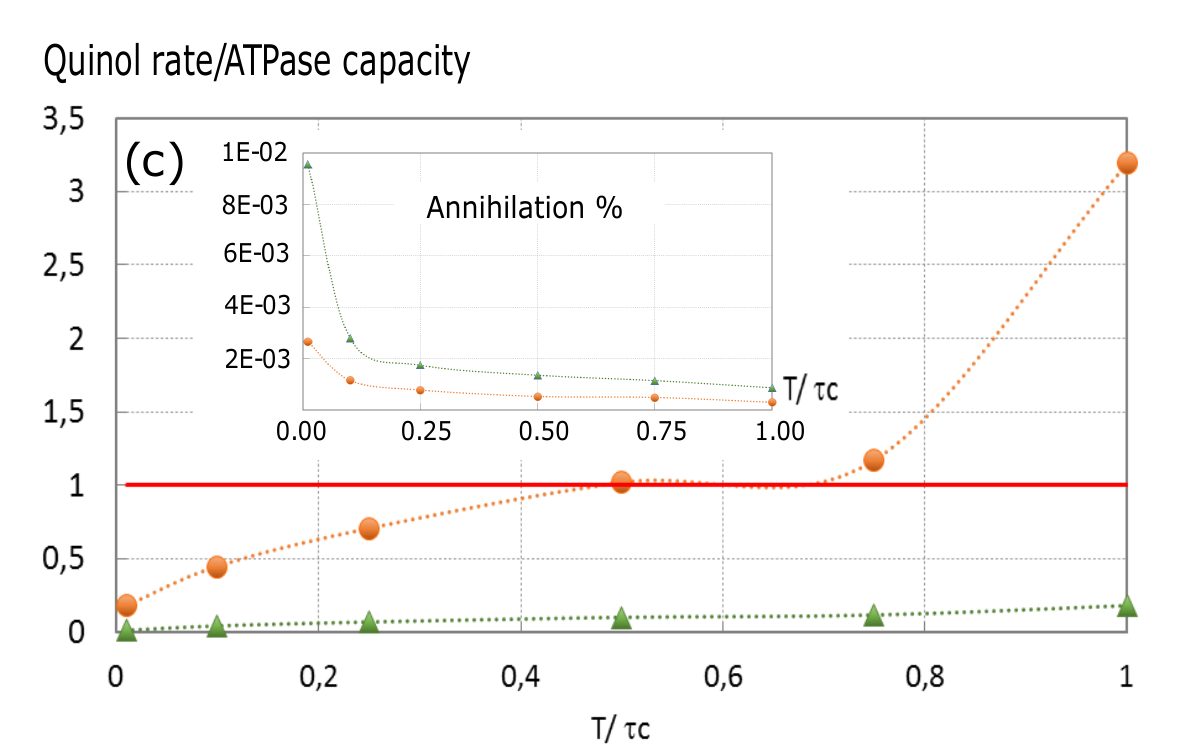}
		\caption{\textbf{Effect of thermal light correlations on quinol production.}
(a) Burstiness $B$ of the absorption traces and of the average burstiness of excitations reaching each RC. In (b) Quinol yield $\eta$, for different temporal correlation $T/\tau_c$ of the absorbed excitations. (c) Quinol production rate, relative to the maximum capacity of the ATPase (horizontal continuous line) as a function of the temporal correlation $T/\tau_c$. Inset: ratio of annihilation events  as a function of the temporal correlations ($T/\tau_c$). $t_{crit}=3000$ ms and $10^{6}$ absorption events.  In (a) and (b) we use high intensity $\langle I\rangle=10^{-3}$photons/(LH$\cdot$ms), while in (c), the results for this intensity (circles) are shown along  the results for low intensity  $\langle I\rangle=10^{-4}$photons/(LH$\cdot$ms) (triangles).}
		\label{fig3}
   \end{figure}

A crucial observation of this work is that the bursted structure of thermal field's statistics is preserved for excitations reaching the RCs. This happens  because the transfer of excitons to the reaction center happens in a time scale ($\sim$ps) that is much shorter than the time between bursts ($\sim$s).  In detail, the calculation of the burstiness $B$ for the simulated waiting times between absorbed excitations, and for the simulated waiting times between excitations reaching each of the RCs, shows that the bursts characteristic of thermal light drive a bursted arrival of excitations to individual RCs, as shown in Fig.\ref{fig3}(a). Since the charge recombination depends on a pairwise exciton arrival to the RC, it might be sensitive to the burstiness of the initial absorption and therefore quinol production may also be influenced by the arrival statistics.

In Fig.\ref{fig3}(b) we show how temporal correlations of thermal light can tune the quinol production, as quantified by the quinol yield $\eta=2\dot{N}_{QH_2}/I_{tot}$, where $\dot{N}_{QH_2}$  represents the rate of quinol production  in the stationary state and $I_{tot}$ is the   photon absorption rate of the simulated photosynthetic vesicle. Notice that the quinol yield decreases for higher temporal correlations of the absorbed photons.
Thermal light exhibits long waiting time intervals  $\langle t_{inter}\rangle$ which allow the relaxation of the $P^+Q_B^-$. In this way, higher correlations in the absorption events result in lower efficiency of quinol production. Temporal correlations arise whenever spatial correlations of thermal light are appreciable. Although spatial correlations are high on the length-scale of bacterial vesicles (cf. Fig \ref{photos}), their almost constant spatial profile across a full vesicle explains the robust performance of vesicles to changes in the specific geometrical arrangements of  LH1 and LH2s (see Appendix C for details).

Under average physiological light intensity $\langle I\rangle=10^{-3}$photons/(LH$\cdot$ms), which is near to the maximum ATPase capacity, and recombination lifetime $t_{crit}$ between 100 and 1000 ms, the quinol efficiency of the membrane varies across the full range $\simeq 0-90\%$, depending on the ratio $T/\tau_c$ (see Fig. \ref{fig3}(b)). For this light intensity, we obtain that the efficiency is {\it strongly} dependent on the degree of temporal correlation from absorbed photons. The quinol yield is kept on very small levels for increasing values of $t_{crit}$ the lower the burstiness, as shown in Fig.\ref{fig3}(b). Moreover, Fig. \ref{fig3}(c) depicts the quinol production rate relative to the ATPase maximum capacity. For physiological light intensity, this figure shows an interesting crossover for $T \lesssim 0.7 \tau_c$  where  the maximum ATP turnover  is achieved (the exact crossover depends on the particular value of $t_{crit}$). This last observation identifies a possible photoprotection mechanism. For the same membrane vesicle  under much lower light intensity $\langle I\rangle=10^{-4}$ photons/LH$\cdot$ms, this maximum turnover is never reached. In this situation a very low $\simeq 0\%$ quinol yield is due to the recombination mechanism, independent on the degree of temporal correlations in the absorbed light. The lifetime of charge recombination within the RCs seems to place an important constraint on the physiological light intensity for the survival of photosynthetic bacteria. The sensitivity of a metastable pairwise charge separation  to thermal light correlations can therefore help tune quinol production to the ATPase turnover capacity in order to avoid excess acidity and cell damage. 

The strong photo-protection mechanism is a consequence of the metastable state lifetime and delayed arrival of bursts, and does not result from additional dissipative processes like annihilation. 
When a burst of photons arrives to the membrane, two photo-excitations have a higher probability to coincide in the same LH structure and annihilate \cite{refannihil}. The dissipation by annihilation, shown in the inset of Fig. \ref{fig3}(c) increases with temporal correlations for the physiological intensity, but its effect remains minimal (a decrease in quinol yield of  $\simeq 0.1-1\%$) in contrast to the decrease in performance due to the temporal correlations  of photo-excitation events shown in Fig.\ref{fig3}(b).\\	

\section{Discussion}

This work studies the role of the statistics of absorption events in the performance of photosynthesis. In our model we have shown that the performance of photosynthesis is affected by the  spatio-temporal correlations present in thermal sunlight.  The interplay between these correlations and the dynamics within RCs,  sets constraints to the light intensities appropriate for bacterial survival, while it provides insight into a pathway for photo-protection, which balances the long intervals between thermal light bursts, and charge recombination taking place in the RCs.
Our work underline that not only the average light intensity and the exciton dynamics, but also the  statistics of the absorption events and the RC charge dynamics, are relevant aspects for the versatile performance of photosynthetic membranes.

We note that this photo-protective mechanism might operate in different ecological niches, since it encompasses the properties of the exciting field common to all forms of photosynthesis  (thermal light), and the dynamics on the RCs which is, except for minor changes,  conserved across species. This may suggest a concept of evolutionary fitness of pairwise charge separation for quinol production and synthesis of ATP, which has endured due to the spatio-temporal correlation present in thermal light.  Arguably, the actual quantification of the statistics of absorption  by a finite absorption linewidth requires further development and provides an interesting perspective for future work, but we should stress that beyond this specific quantification, the presented unified analysis of absorption, excitonic transfer and charge dynamics already provides a glimpse on a regulation mechanism of biological photosynthesis that is driven by the coherence of thermal sunlight.

Besides opening new avenues for further theoretical studies, we stress that our findings lead to predictions that can be experimentally tested with present resources to measure cellular fitness and regulation, e.g., bacterial growth rate, metabolic activity, or gene expression among others. Comparison of fitness indicators for bacteria grown under illumination with attenuated laser and pseudo-thermal sources \cite{Haner,Valencia} will shed light on the response of these organisms to the spatio-temporal correlations of the incident light. Better fitness indicators for pseudo-thermal light will confirm the significance of the interplay of light absorption statistics and charge recombination, in agreement with our predictions.

Moreover, the statistics of illumination have been shown to improve the {\it rectification} stage, i.e. Alternating-to-Direct current conversion in solar antennas \cite{Mashaal1,SAreview,Boriskina}. Rectification nowadays drops the solar antennae collection efficiency ($>80\%$) to about $0.01\%$ \cite{Vandenbosch, Zhao}, but might  be improved by the
coherent combination of the light detected by different antennas, i.e., the spatio-temporal correlations \cite{Mashaal1,SAreview,Boriskina}.  In view of our results, it is an interesting perspective to understand to what extent the spatio-temporal correlations of thermal light could also assist the rectification process in artificial light-harvesting technologies.\\

\section{Methods}

\subsection{Statistical model of thermal light detection.}

The absorption of light by a set of receptors is a stochastic process, in which the absorption events are not independent of each other. The quantum photon-interference leads to spatial and temporal correlation between absorption events, affecting the respective photon-counting statistics \cite{Scully}. Our aim is to introduce the spatio-temporal correlations into the calculation of the probability $f(t)$ for the time elapsed between consecutive detections $t$. This is used to generate the photon-absorption events required at the beginning of our photosynthesis simulations. That is, our algorithm generates a random number between 0 and 1 for $F(t)=\int_0^t f(t') dt'$ and resolves the corresponding $t$ value, which is the elapsed time since the last detection event until the current one.\\

In summary the calculation workflow reads as follow: The spatial $\gamma_{k,l}$ and temporal $\gamma(t_1,t_2)$ normalized correlation functions are plugged into the generating functional $G(s,T)$, which in turn allows the calculation of the cumulative probability distribution for inter-event time $F(t)$. Finally $F(t)$ is used to generate the photon-absorption events that will be subsequently converted into photo-excitations in the Montecarlo simulations. Schematically, this is:

\begin{equation*}
\gamma_{k,l},\gamma(t_1,t_2) \rightarrow G(s,T) \rightarrow F(t) \rightarrow \text{ simulations}
\end{equation*}

For the convenience of the reader, we describe below all the quantities involved, while we refer to the Appendix A1 for a review of their derivation for the relevant field; namely a gaussian light field in the far-field approximation, assuming a low absorption bandwidth.\\ 

The calculation starts with the spatio-temporal normalized correlation functions (also called coherence functions) $\gamma_{k,l}(t_1,t_2)=\left\langle \hat{E}^+(\vec{r}_k,t_1)\hat{E}^-(\vec{r}_l,t_2)\right\rangle/\sqrt{\left\langle n_k\right\rangle}\sqrt{\left\langle n_l\right\rangle}$. Here $\hat{E}^{\pm}$ is the electric field operator, $k$,$l$ refer to a couple of absorbers, $r_{k(l)}$ is the position of the detector labelled $k$($l$), which receive photons at time $t_{1(2)}$; and $\langle n_i \rangle$ is the average number of photon-absorptions at each detector. Assuming that the superposition of the light field coming from different points at the source, does not affect its spectral properties at the reception positions -- a feature called \textit{cross-spectral purity} -- the spatial and temporal contributions can be separated $\Gamma_{k,l}(t_1,t_2)=\sqrt{\left\langle n_k\right\rangle}\sqrt{\left\langle n_l\right\rangle}\gamma_{k,l} \gamma(t_1,t_2)$ \cite{Mandel-CEP}. These coherence functions are obtained for quasi-monochromatic light in the far field approximation \cite{Mandel,Shih} (see Appendix A1 for derivation):

\begin{align}
\label{cfa} \gamma_{k,l}&=\frac{J(\nu_{k,l})}{\nu_{k,l}},\\
\label{cfb} \gamma(t_1,t_2)&=\frac{\sin(\tau_{1,2})}{(\tau_{1,2})}.
\end{align}

The spatial correlations are described by eq. \ref{cfa}, where $J(\nu_{k,l})$ is the first order Bessel function, and the argument $\nu_{k,l}=2\pi\vert \vec{r}_{k,l}\vert/l_c$ compares the inter-detectors distance with the transverse coherence lenght $l_c$. The spatial correlation function $\gamma_{k,l}$ quantifies the decay of the spatial correlations as a function of the distance between the detectors $k$ and $l$. Spatial coherence is relevant for distances inside the transverse coherence length $\vert \vec{r}_{k,l}\vert \lessapprox l_c$. In turn, the transverse coherence length $l_c= \lambda \frac{D}{a}$, relates the average wavelength $\lambda$ of the field, the distance to the light source $D$, and  its diameter $a$, as depicted in Fig.\ref{photos}(a) \cite{ADM-IOP}. Within the coherence area, temporal correlations of thermal light are accounted for by a temporal correlation function (eq. \ref{cfb}), whose argument for a stationary field is $\tau_{1,2}=\frac{2\pi(t_2-t_1)}{\tau_c}$.  Similarly to spatial case, $\gamma(t_1,t_2)$ describes the decay of the correlations as a function of the inter-detections time $(t_1-t_2)$, normalized by the coherence time of the light $\tau_c$. The maximum possible value that $(t_1-t_2)$ can get is the detection time $T$. These temporal and spatial normalized correlation functions have been experimentally well tested, see references \cite{Shih1,Shih2}.\\

Correlation functions (eqs. \ref{cfa} and \ref{cfb}) are introduced into the statistical description of the light-absorption process by means of
 the \textit{factorial moments generating function} $G(\{s_i\},T)$, which encodes all the statistical information of the process \cite{Zardecki,Bures72,Van_Kampen,Bures71}. This formalism accounts for the statistics of the photodetection events recorded by a set of $N$ detectors, each of which we identify as individual LH units in our case. Here we use the generating functional formalism to calculate the photocounting joint probability $P(n_1,n_2,...,n_N,T)$, that $n_i$ absorption events occur at the $i$-th LH ($i=1,...,N$) during the time window $T$ \cite{Bedard,Zardecki,ADM-IOP,Bures71,Bures72}:

\vspace{-0.2cm}
\begin{equation}
P(n_1,n_2,...,n_N;T)=
\left \{ \prod_{i=1}^N \frac{(-1)^{n_i}}{n_i!}\frac{\partial^{n_i}}{\partial s_i^{n_i}}  \right \}G(\{s_i\},T)|_{\{s_i=1\}} \text{ .}
\label{Eq:a1}
\end{equation}

The probability to absorb $n$ photons in the total set of detectors, regardless of the specific counting record of any individual detector, is then

\begin{equation}
P(n,T)=\sum_{\{ n_i \}}\delta \left ( n-\sum_{i=1}^Nn_i\right )P(n_1,n_2,...,n_N,T).
\label{ptotal}
\end{equation}

The calculation of the generating functional for Gaussian light leads to \cite{Zardecki,Bures72}:

\begin{equation}
G(\left\{s_i\right\},T)=\prod_{j=1}^T \prod^{N}_{k=1} (1+s_i\varpi_jb_k)^{-1} \text{ ,}
\label{G_gen}
\end{equation}

\noindent such that the spatial and temporal coherence functions enter into the model by means of the eigenvalues $\varpi_j$ and $b_k$ of the spatial and temporal Fredholm equations (see Appendix A2 for the derivation),

\begin{align*}
\int^T_0 dt_2 \gamma(t_1-t_2) \psi_j(t_2)&=\varpi_j \psi_j(t_1)\\
\sum_{l=1}^N \sqrt{\alpha_k \langle n_k\rangle}\gamma(r_{k,l})\sqrt{\alpha_l \langle n_l\rangle}\phi_l(t)&=b_k \phi_k(t) \text{.}
\end{align*}

Finally, it can be demonstrated \cite{Rockower} that the cumulative distribution function $F(t)=1-P(n=0,t)$ (see Appendix A3 for calculation). Therefore, $F(t)$ just depends on the generating functional $G(s,T)\vert_{s=1}$, which in turn depends on the aperture time $T$ (time window for photo-detection), and on the  temporal and spatial normalised correlation functions \cite{ADM-IOP}. \\

This photo-detection theory was developed for avalanche photo-diodes where the detection time is a controllable experimental parameter. Therefore the main limitation of this formalism to describe an actual LH complex is the definition of this time window $T$ for photon-arrivals counting. Despite the large amount of existent works regarding the theory of photon-statistics of finite band-width light absorption \cite{Glauber_1963PR,Mandel_1964PPS,Rosseau_1975,Rosseau_1977,Kimble}, a fully comprehensive theory applicable to chromophores is yet to be developed. 
Notwithstanding, the ratio $T/\tau_c\ll 1$ or $T/\tau_c\gg 1$, sets the limits of Bose-Einstein (maximum correlations) or Poissonian (no correlations) statistics for light absorption, respectively.
For  $T/\tau_c\lesssim 1$, this  framework results in waiting time distributions $f(t)$ that exhibit longer tails than the exponential distribution expected from independent events \cite{Rockower}. The consequence of these slow-decaying tails is a bursted  structure \cite{Barabasi}, with photo-excitation traces that have very long waiting times scattered between bursts of clustered events (typical traces shown in Fig.\ref{incoming}(a). \\

We simulate a full photosynthetic vesicle depicted in Fig.\ref{Trates1}(b), consisting of 400 LHs (44 core complexes and 356 LH2 antenna complexes), in agreement with realistic stoichiometry for the simulated light intensity \cite{Scheuring}. Simulations were performed for two fixed light intensities ($\langle I\rangle=10^{-3}$photons/(LH$\cdot$ms) or $\langle I\rangle=10^{-4}$photons/(LH$\cdot$ms)), scanning different values of $T/\tau_c$ always verifying that the average rate of photon arrivals is preserved and equals the light intensity. \\

Although equation \ref{Eq:a1} allows to obtain the probability of photon arrivals at each LH, the calculation of the photo-detection traces for the total set of 400 LH units is a very expensive computational task. Instead of that, we simulate the photo-detection traces for the full membrane (eq. \ref{ptotal}) and in a next step we spatially distribute the detected photons in the membrane according to the second-order spatial coherence function of the intensity (see Appendix C for details). 
 \\

\subsection{Photosynthesis simulations}


The generated photo-excitation traces are coupled to a dynamical model of the photosynthetic vesicle, in which the excitons travel through the membrane by individual hopping processes between LH rings until reaching an available RC. This transport process is based upon experimental estimations of excitonic transfer rates and charge transfer dynamics \cite{Caycedo_2010PRL}.  Here we also include the recombination of the intermediate metastable state $P^+Q_B^-$, which is described by a stochastic process with associated lifetime $t_{crit}$ after  charge separation $P^*\rightarrow P^+$ occurred.

\begin{table}
\begin{center}
\begin{tabular}{|c |c |c |}
\hline
\textbf{Process} & \textbf{Value} & \textbf{Data source [Reference]} \\
\hline
LH2$\rightarrow$LH2 &  $t_{22}=10$ps &	Calculated \cite{review}\\
LH2$\rightarrow$ LH1 & $t_{21}=3.3$ps &	Measured for {\it R. Sphaeroides} \cite{Hess}\\
LH1$\rightarrow$ LH2 & $t_{12}=15.5$ps & Measured for {\it R. Sphaeroides} \cite{Hess}\\
LH1$\rightarrow$ LH1 & $t_{11}=20.0$ps &  Calculated by F\"orster interaction \cite{ritz}\\
LH1$\rightarrow$ RC  & $t_{1,RC}=25$ps &  Calculated at $300$ K \cite{Vgrondelle}\\
RC$\rightarrow$ LH1  & $t_{RC,1}=8.0$ps & Calculated \cite{Damja} and measured \cite{Timpmann}\\
Dissipation & $\gamma_{D}=1$ns$^{-1}$ & Estimated from Fluorescence \cite{ritz}\\
 \hline
\end{tabular}
\caption{\footnotesize{Considered excitonic transfer rates $t_{ij}$ ($i,j=1,2$,RC) for the simulation of the primary steps of photosynthesis in purple bacteria.}}
\end{center}
\label{Trates}
\end{table}

After a detection event is simulated, the photo-excitation is located on the corresponding ring (LH1 or LH2), and travels to a reaction center with some probability of being dissipated during the transference. Once the excitation has arrived at the RC, the latter performs the reduction-oxidation cycle described before, and two of these photo-excitations are necessary to form a quinol molecule ($Q_{B}H_{2}$). In the simulations, when a RC receives the first photo-excitation there is a check for the metastable condition; if the next one does not arrive during the life-time window $t_{crit}$, the charge is recombined and the exciton is wasted (in this case the next photo-excitation will occupy its place). Otherwise, the reaction center is closed and not available during some time $\tau_{RC}=10$ms while finishing the process and produces a quinol molecule \cite{Caycedo_Soler_2010NJP}. Although the turnover of the RCs depends on the acidity, we did not include that dependence and only chose an intermediate rate of quinol production (100 per second at pH $~ 5$ \cite{Osvath}), in such a way that the closing time of the reaction centers obeys an exponential distribution with that average throughput. The transfer rates used in our simulations are summarized in Table I, with values taken from pump-probe experiments or F\"oster calculations assuming intra-complex delocalization, which is in good agreement with experiments \cite{review,vanGrondelle,Schroeder}. \\

In the photosynthetic membrane, after a quinol is produced, its two hydrogens are dissociated and used by the ATP-ase enzime to produce  ATP. It is worth to mention that as little as 5 RCs already saturate the capacity of the ATP-ase, in the case of slowest RCs turnover equal to 50 quinols per second for a pH = 2 \cite{Osvath}. 
The mismatch between the overall quinol production and the ATPase turnover determines the bottleneck of the system, restricting the maximum quinol throughput that can be converted into ATP and the incoming-photons rate to 200 photons/s. In our simulations, we use the average intensity 0.4 photons/ms for the whole vesicle ($\left\langle I\right\rangle =10^{-3}$photons/ms$\cdot$LH), which is processed without limitations by the set of 44 RCs depicted in fig.\ref{Trates1}(b), whose minimum total capacity is 4.4 photons/ms at pH=2 \cite{Osvath}.\\


\textbf{Acknowledgements} This work was supported by the ERC Synergy grant BioQ, and by COLCIENCIAS and Universidad de Los Andes. We thank Neil Johnson, Pedro Manrique, Dario Egloff and A.D,M,'s thesis supervisors Luis Quiroga and Ferney Rodriguez for very interesting discussions and comments. This publication was made possible through the support of the John Templeton Foundation.

\appendix

\section{Calculations to obtain the coherence functions, the generating functional and the probability distribution of the inter-absorptions time}

\subsection{Spatial and temporal coherence functions}

In this section we review the derivation of the second order correlation function for a thermal field in the far-field approximation for small absorption bandwidth. The correlation function is defined as the mean value of the product of the fields at different positions and times \cite{Mandel,Scully,Shih1,Shih2,Shih3}:

\begin{equation}
\Gamma_{i,j}(t_1,t_2)=\left\langle \hat{\vec{E}}^{(-)}(\vec{r}_i,t_1) \hat{\vec{E}}^{(+)}(\vec{r}_j,t_2) \right\rangle,
\label{corrdef}
\end{equation}

\noindent where $\left\langle \ldots \right\rangle$ means the average over the system states. $\vec{E}^{(\pm)}(\vec{r}_i,t_i)$ are respectively the positive and negative electric field operators, defined by

\begin{align}\label{ints}
\nonumber \hat{\vec{E}}^{(+)}(\vec{r}_i,t_i)&=\sum_\mu \int d{\vec{k}} \varepsilon_k g_{\vec{k},i} \hat{a}_{\mu,\vec{k}},\\
\hat{\vec{E}}^{(-)}(\vec{r}_i,t_i)&=\sum_\mu \int d{\vec{k}} \varepsilon_k g^*_{\vec{k},i} \hat{a}_{\mu,\vec{k}}^{\dagger}\text{ .}
\end{align}

\noindent Here the index $\mu$ accounts for polarization, $\varepsilon_k=\sqrt{\frac{\hbar ck}{2\epsilon_0 (2\pi)^3}}$, with $c$ being the speed of the light, $k=|\vec{k}|$ the magnitude of the wave vector, and $\epsilon_0$ the electric permittivity. The creation and annihilation operators of photons in the respective mode are $\hat{a}_{\mu,\vec{k}}^{\dagger}$ and $\hat{a}_{\mu,\vec{k}}$. The Green function $g_{\vec{k},i}$ that propagates the longitudinal and transverse light modes from the source (in the origin) to the detection point at $(\vec{r}_i,t_i)$ is 
$g_{\vec{k},i}=e^{i(\vec{k}\cdot \vec{r}_i-\omega t_i)}$, with $\omega=ck$. Introducing eqs. \ref{ints} into eq. \ref{corrdef}, the second order correlation of the field, we get: 

\begin{align}
\nonumber \Gamma_{i,j}(t_1,t_2) &= \frac{\hbar c}{2\epsilon_0(2\pi)^3} \int \int d\vec{k}d\vec{k'} \eta_d(k,k') \sqrt{k k'}e^{i(\vec{k}\cdot \vec{r}_i-\omega t_1)-i(\vec{k'}\cdot \vec{r}_j-\omega' t_2)} \\
&\hspace{4cm} \times \left\langle \hat{a}_{\vec{k}}^{\dagger} \hat{a}_{\vec{k'}}\right\rangle.
\end{align}

In the last equation polarization has been ignored because it only affects the normalization factor (cf. eq.10.8-2 of reference \cite{Mandel}). Notice that we have introduced a density function $\eta_d(k,k')=(H(k_0+\Delta k)-H(k_0-\Delta k))(H(k'_0+\Delta k')-H(k'_0-\Delta k'))$ for finite bandwidth absorption in the range $\Delta k$ around the central wave number $k_0$, where $H(k)$ is the heaviside step function. Bosonic thermal correlation is included through $\left\langle \hat{a}_{\vec{k}}^{\dagger} \hat{a}_{\vec{k'}}\right\rangle=\delta_{\vec{k},\vec{k'}} n(k,\mathbb{T})$, where $n(k,\mathbb{T})=\frac{1}{e^{\hbar ck/k_B \mathbb{T}}-1}$ is the number of photons with frequency $\omega=ck$, which come from a thermal source at thermal equilibrium with a reservoir at temperature $\mathbb{T}$. Replacing into the last equation, the correlation function is

\begin{equation}\label{int2}
\Gamma_{i,j}(t_1,t_2)=\frac{\hbar c}{2\epsilon_0(2\pi)^3} \int d\vec{k} e^{i(\vec{k}\cdot \vec{r}_{i,j}-\omega t_{1,2})}\frac{k}{e^{\hbar c k/k_B \mathbb{T}}-1},
\end{equation}

\noindent where we have replaced $\vec{r}_{i,j}=\vec{r}_i-\vec{r}_j$ and $t_{1,2}=t_1-t_2$. The integral in eq. \ref{int2} can not be evaluated in a closed form, but reducing the number of detected modes, by assuming small absorption bandwidth and far field detection, a closed expression can be obtained. This approximations are particularly useful in the case of Sunlight absorption on Earth, where light-harvesting complexes absorb on a sharp interval of the light spectra. Lets calculate the spatial case first. In the far field approximation the product $\vec{k}\cdot \vec{r}_{i,j}$ is written in spherical coordinates: 

\hspace{-0.3cm}
\begin{equation}
\vec{k}\cdot \vec{r}_{i,j}=k\; r_{i,j}\sin(\theta_k)\cos(\phi_{r_{i,j}}-\phi_k),
\end{equation}

\noindent where $\phi$ and $\theta$ are the polar and azimuthal angles for the vector indicated in the sub-indexes. Here the azimuth angle has been set to $\theta_{r_{i,j}}=\pi/2$ for a flat detection surface, and  $\theta_s$ is the azimuthal angle covered by the source. In the far field approximation the angle $\theta_k$ is taken to be small but not null, allowing to replace $\sin(\theta_k)\rightarrow \theta_k$, then the correlation function can be written:

\begin{align}
\Gamma_{i,j}(t_1,t_2)&=\frac{\hbar c}{2\epsilon_0(2\pi)^3} \int d\vec{k} (H(k_0+\Delta k)-H(k_0-\Delta k)) \\
\nonumber  & \hspace{2.5cm} \times \frac{ke^{ik\; r_{i,j}\theta_k \cos(\phi_{r_{i,j}}-\phi_k)}e^{-i\omega t_{1,2}}}{e^{\hbar c k/k_B \mathbb{T}}-1} \\
\label{corrdefs2} &=\frac{\hbar c}{2\epsilon_0(2\pi)^3} \int_{\Delta k} dk \; \frac{k^3 e^{-i\omega t_{1,2}}}{e^{\hbar c k/k_B \mathbb{T}}-1} \\
\nonumber &\hspace{2cm} \times \int_0^{\theta_s}\vspace{-0.4cm}\int_0^{2\pi} \vspace{-0.4cm} d\theta_k \; d\phi \; \theta_k e^{ik r_{i,j}\theta_k\cos(\phi_{i,j}-\phi_k)}\\
\label{corrdefs3} &=\frac{\hbar}{2\epsilon_0(2\pi c)^3} \int_{\Delta \omega} d\omega \; \frac{\omega^3 e^{-i\omega t_{1,2}}}{e^{\hbar \omega/k_B \mathbb{T}}-1} \left[\frac{2\pi \theta^2_s J_1(\theta_s r_{i,j}\omega/c)}{(\theta_s r_{i,j}\omega/c)}\right],
\end{align}

\noindent where we made use of the dispersion relation $\omega=ck$. The angular integral in eq.\ref{corrdefs2} is well known in the theory of Fraunhofer diffraction for a circular aperture \cite{Born}, with $J_1(\ldots)$ being the first order Bessel function. Now we normalize the correlation function to write

\begin{align}
\nonumber \hspace{-1cm} \gamma_{i,j}(t_1,t_2)&=\frac{\left\langle \hat{\vec{E}}^{(-)}(\vec{r}_i,t) \hat{\vec{E}}^{(+)}(\vec{r}_j,t)\right\rangle}{\left\langle \hat{\vec{E}}^{(-)}(\vec{r},t) \hat{\vec{E}}^{(+)}(\vec{r},t) \right\rangle} \\
\label{gst1} &=\frac{\int_{\Delta \omega} d\omega \; \frac{\omega^3 e^{-i\omega t_{1,2}}}{e^{\hbar \omega/k_B \mathbb{T}}-1} \left[\frac{ J_1(\theta_s r_{i,j}\omega/c)}{(\theta_s r_{i,j}\omega/c)}\right]}{\int_{\Delta \omega} d\omega \; \frac{\omega^3}{e^{\hbar \omega/k_B \mathbb{T}}-1}}.
\end{align}

Assuming cross-spectral purity \cite{Mandel-CEP}, the spatial correlations are not affected by the spectral properties of the field, and can be taken out of the integral (evaluated at a mean  frequency $\omega_0$)

\begin{align}
\nonumber \hspace{-1cm} \gamma_{i,j}(t_1,t_2)=\left[\frac{ J_1(\theta_s r_{i,j}\omega_0/c)}{(\theta_s r_{i,j}\omega_0/c)}\right]\frac{\int_{\Delta \omega} d\omega \; \frac{\omega^3 }{e^{\hbar \omega/k_B \mathbb{T}}-1}e^{-i\omega t_{1,2}}}{\int_{\Delta \omega} d\omega \; \frac{\omega^3}{e^{\hbar \omega/k_B \mathbb{T}}-1}}.
\end{align}

Now, applying the limit of small detection bandwidth ($\Delta \omega \rightarrow 0$), the factor $\frac{\omega^3 }{e^{\hbar \omega/k_B \mathbb{T}}-1}$ can be taken out of the integral. In the range of finite absorption bandwidth by LH complexes ($\lambda \approx 800 \pm 50$nm), this factor changes $\sim \pm$1.4\%. After this step the last equation reads

\begin{align}
\nonumber \hspace{-1cm} \gamma_{i,j}(t_1,t_2)=\left[\frac{ J_1(\theta_s r_{i,j}\omega_0/c)}{(\theta_s r_{i,j}\omega_0/c)}\right]\frac{1}{\Delta \omega}\int_{\Delta \omega} d\omega \; e^{-i\omega t_{1,2}},
\end{align}

\noindent which (up to a global phase) is 

\begin{equation}
\label{gamma}\gamma_{i,j}(t_1,t_2) \approx \sinc\left(\tau_{1,2}\right)\frac{J_1\left(\nu_{i,j}\right)}{\left(\nu_{i,j}\right)}.
\end{equation}

\noindent Here the arguments are $\nu_{i,j}=\theta_s r_{i,j}\omega/c=2\pi\vert \vec{r}_i-\vec{r}_j\vert \mathtt{a}/\lambda D$ and $\tau_{1,2}=\Delta \omega(t_2-t_1)$, where we have replaced $\theta_s \approx \mathtt{a}/D$ (valid for small angles); that is the ratio between the radius of the source $\mathtt{a}$ and the distance between the source and the reception surface $D$.

\subsection{Factorial moments generating functional of thermal light detection}

In general, light has correlations at different orders. Therefore, to describe light detection it is appropriate to use the \textit{factorial cumulants generating functional} $C(\left\{s_i\right\},T)$, which is defined as the infinite series \cite{Bures71,Van_Kampen}

\begin{align}
\nonumber C(\left\{s_i\right\},T)&=\ln G(\left\{s_i\right\},T)\\
&=\sum_{m=1}^{\infty} \frac{(-1)^m(s_1s_2\ldots s_m)}{m!}k_{[m]},
\label{C}
\end{align}
\vspace{-0.2cm}

\noindent where $\lbrace s_i \rbrace$ is a set of expansion parameters, and $k_{[m]}=(m-1)!(\alpha_1 \alpha_2\ldots \alpha_m) \Tr\left\{ \Gamma^{(m)}\right\}$ is the m-th order factorial cumulant, enclosing the $m$th order correlation ($\Gamma^{(m)}$). Here $\alpha_i$ is the quantum efficiency of the $i$th detector\footnote{The definition of $C(s)$ contains an infinite series which allows to describe the all-order correlations in the stochastic process, but $N$ detectors can maximally detect N-order correlations} and $ \Tr\left\{ \Gamma^{(m)}\right\} \equiv \sum^{N}_{l=1} \int^T_{0} dt \Gamma^{(m)}_{l,l}(t,t)$. The $m$th order correlation function $\Gamma^{(m)}$ can be expressed in terms of lower order correlations, see \cite{Bures72,Cantrell}:
\vspace{-0.2cm}

\begin{align}\label{ring}
\Gamma^{(m)}_{k,l}&(t_1,t_2)=\sum^{N}_{i=1}\int^T_0 dt \Gamma^{(m-1)}_{k,i}(t_1,t)\Gamma^{(1)}_{i,l}(t,t_2) \hspace{0.4cm}  \text{$m\geq 2$ }.
\end{align}
\vspace{-0.2cm}

The second order spatio-temporal correlations of the field (defined in the previous section) are also called the first order intensity correlations, since in eq. \ref{ring} $\Gamma^{(1)}_{k,l}(t_1,t_2)=\left\langle \hat{E}^+(\vec{r}_k,t_1)\hat{E}^-(\vec{r}_l,t_2)\right\rangle$. For simplicity, we will write from now on $\Gamma_{k,l}(t_1-t_2)$ for the stationary field.  Including all of these in Eq.\ref{C}, the factorial moments generating functional gets:

\begin{align}\label{Gs1}
\nonumber G&(\left\{s_i\right\},T)=\\
\nonumber &\exp{\left\{\sum^{N}_{l=1} \int^T_{0}  dt \sum_{m=1}^{\infty} \frac{(-1)^m}{m}(s_1 s_2\ldots s_m)(\alpha_1 \alpha_2\ldots \alpha_m) \Gamma^{(m)}_{l,l}(t,t) \right\}} \\
&=\exp\left\{ \Tr\left\{-\left[\ln (\mathbb{D})\right]\right\}  \right\}=\frac{1}{\det\left[\mathbb{D}\right]};
\end{align}

\noindent where we have defined the matrix $\mathbb{D}=\textbf{1}+\sqrt[m]{(s_1 s_2\ldots s_m)(\alpha_1 \alpha_2\ldots \alpha_m)} \Gamma^{(m)}$ and $\Gamma^{(m)}$ includes spatial and temporal correlations \cite{Zardecki}.  Because of the Gaussian moments theorem, the $\mathbb{D}$ matrix for a Gaussian process will be a four-dimensional operator to include second order spatial and temporal field correlations:
\vspace{-0.2cm}

\begin{equation}
\mathbb{D}_{k,l}(t_1-t_2)=\delta_{k,l}\delta(t_1-t_2)+\sqrt{s_k \alpha_k} \;\Gamma_{k,l}(t_1-t_2)\sqrt{s_l \alpha_l},
\label{Dkl}
\end{equation}

\noindent to reach
\vspace{-0.2cm}

\begin{equation}
G(\left\{s_i\right\},T)=\frac{1}{\det\left[\mathbb{D}^*_{k,l}(t_1-t_2)\right]}=\frac{1}{\prod_i\kappa_i}.
\label{GF2}
\end{equation}

\noindent Here $\left\lbrace  \kappa_i \right\rbrace$ is the set of eigenvalues  of the integral equation \
 $\sum_{l=1}\int^T_0 dt_2 \mathbb{D}^*_{k,l}(t_1-t_2) \Phi^{(i)}_l(t_2)=\kappa_i\Phi^{(i)}_k(t_1)$. Cross-spectral purity permits to assume factorizable spatio-temporal correlations $\Gamma_{k,l}(t_1-t_2)=\sqrt{\left\langle n_k\right\rangle}\sqrt{\left\langle n_l\right\rangle}\gamma(r_{k,l})\gamma(t_1-t_2)$ into spatial $\gamma(r_{k,l})$ and temporal $\gamma(t_1-t_2)$ correlation functions (defined in the methods section of the main text). The corresponding eigenvalues get $\kappa_i=1+s_i\varpi_{j}b_k$, which depends on the indexes $j,k$ when separating into temporal and spatial equations,
 
\begin{align}
\label{te-eq}	\int^T_0 dt_2 \gamma(t_1-t_2) \psi_j(t_2)&=\varpi_j \psi_j(t_1)\\
\nonumber \text{ and}\\
\label{te-eq1}	\sum_{l=1}^N \sqrt{\alpha_k \langle n_k\rangle}\gamma(r_{k,l})\sqrt{\alpha_l \langle n_l\rangle}\phi_l(t)&=b_k \phi_k(t) \text{ .}
\end{align}

Notice that $\gamma(r_{k,l})$ is the spatial correlation function normalized by the average number of absorbed photons per detector, i.e. each light harvesting ring (LH), such that $\Gamma_{k,l}=\sqrt{\left\langle n_k\right\rangle}\sqrt{\left\langle n_l\right\rangle}\gamma(r_{k,l})$. The average number of photon-absorptions in the $k$-th detector is obtained by $\langle n_k \rangle	=\frac{-dG(\left\{s_i\right\})}{ds_k}\vert_{s_k=0}=\alpha_k N\langle I_k \rangle T$. When $\alpha_k$, $N$ and $T$ are fixed, we will refer to $\langle n \rangle$ or $\langle I \rangle$ as \comillas{intensity} indistinctly. \\ 

In order to have correlated absorption events, the set of LHs should be able to discriminate photons inside the coherence volume, i.e. the spatial extension of the system $l$ to be similar or smaller than the coherence length $l_c$, and the detection time $T$ also similar or smaller than the coherence time $\tau_c$. The spatial extension of the system is set by the maximum inter-detectors distance ($l= \max(|\vec{r}_i-\vec{r}_j|)$). In a similar way, the temporal extension of the absorption is defined in terms of the inter-detections times ($T \sim \text{max}|t_i-t_j|$), where $t_k$ is the time at which the detector $k$ is available to absorb a photon. In the special case of a single absorber, one easily sees that the elapsed time for the detector to be available for two consecutive detections (i.e. the time to distinguish photons) is limited by the recovery time of the absorber. 
In the original experiments with avalanche photo-diodes detectors \cite{Bures71,Bures72,Shih3,Valencia}, the detection time was defined as a binning time, which hence gives the timescale needed to resolve the occurrence time of the absorption events. 
Therefore, $T$ is limited by the time required by the LH complexes to absorb two consecutive photons. Since the capacity of the detection system to distinguish photons inside $\tau_c$ determines its ability to reproduce the correlated structure of light, the ratio  $T/\tau_c$ allows the continuous tuning of the absorbed excitations from Bose-Einstein (maximum correlations at $T/\tau_c\ll 1$) to Poissonian statistics (no correlations at $T/\tau_c\gg 1$).\\

\subsection{Probability distribution for the time elapsed between consecutive photo-detections}

Here we present the procedure described by Rockower in reference \cite{Rockower}, to obtain the probability distribution for a random variable $t$, defined as the time between consecutive photo-detections. We need first to define the quantities used below: $t_{\tilde{n}}$ is the detection time of the $\tilde{n}$-th photon and $n(T)$ is the number of photo-detection events occurring in a time window $T$. If the number of events occurring in T is less than $\tilde{n}$, the $\tilde{n}$-th detection event occurs after $T$:

\begin{equation*}
P(T<t_{\tilde{n}})=P(n(T)<\tilde{n})=\sum^{\tilde{n}-1}_{i=0} P(n=i).
\end{equation*}

On the other hand, there is the cumulative probability function $F(T)=\int^T_0 dt f(t)=P(t_{\tilde{n}}<T)$ and therefore,

\begin{align*}
1-F(T)&=\int^{\infty}_T dt_{\tilde{n}} f(t_{\tilde{n}})\\
&=P(T<t_{\tilde{n}})=\sum^{\tilde{n}-1}_{i=0} P(n=i).
\end{align*}

\noindent Reorganizing, we are left with $F(T)=1-\sum^{\tilde{n}-1}_{i=0} P(n=i)$. We are interested in the time between two successive counts, i.e., the time between the arrival of one photon at the detector and the next one, defined as the \comillas{inter-detection time} $t$. We therefore use the last result with $\tilde{n}=1$ for the time passed before the first detection (cf. \cite{Van_Kampen}eq.6.1):

\begin{align}
\label{Fdt} F(t)&=1-P(n=0,t) \text{ ,}\\
f(t) &=-\frac{dP(n=0,t)}{dt}\text{ ,}
\end{align}

\noindent recalling that the Photocounting probability depends on the detection time itself $P(n)\rightarrow P(n,t)$. Here $f(t)$ is the probability density function for the occurrence of a time-interval $t$ between consecutive detection events.

\section{Thermal light excitation and Charge separation} 

\begin{figure}
			\includegraphics[width=1\columnwidth]{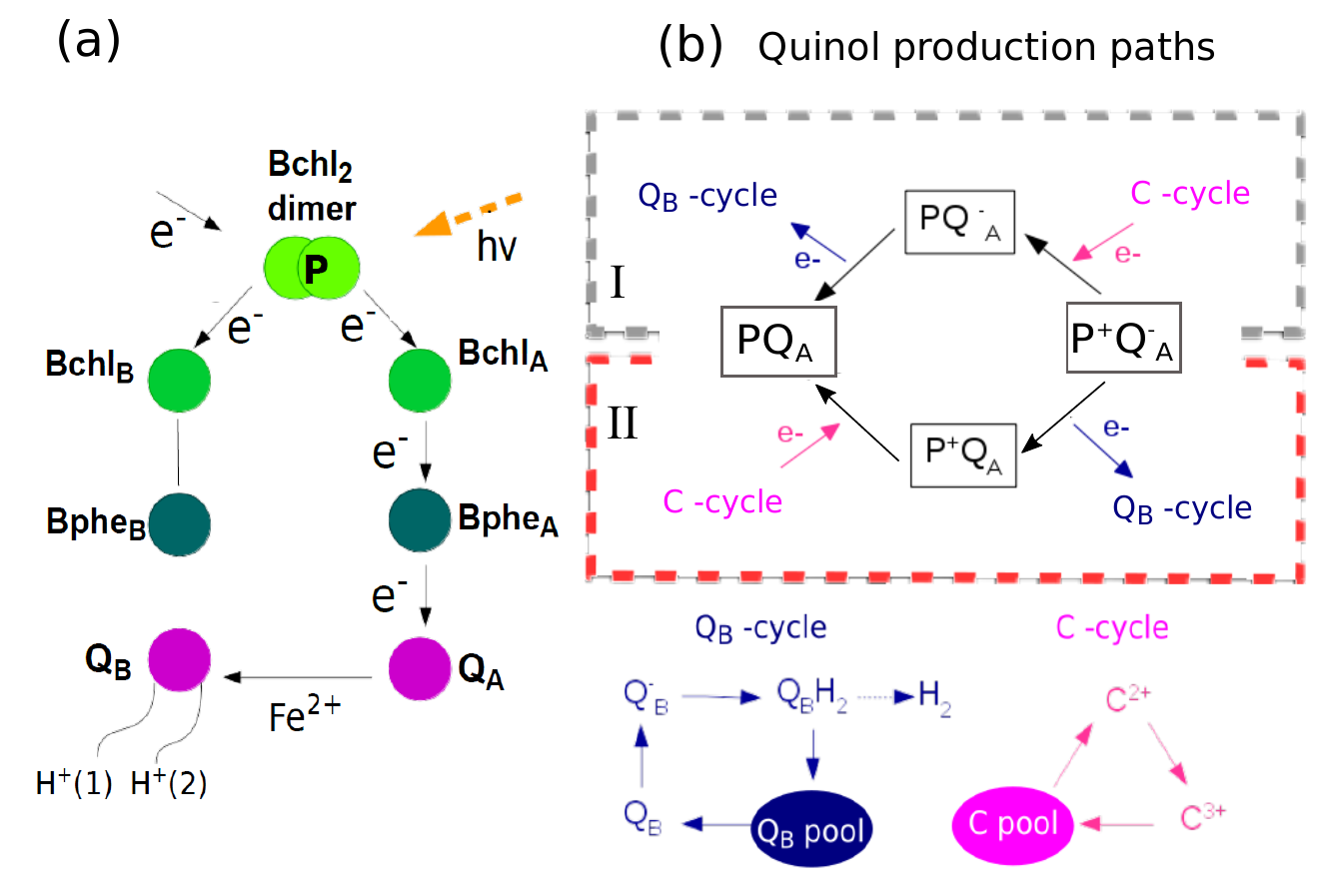}
				\caption{(a)  Scheme for photo-excitation of the special pair  $P$ and subsequent charge transfer  to the accessory pigment $BChl_A$, the bacteriopheophytin ($Bphe_A$) and the quinone  $Q_A$,which take place in the active branch $BChl_A\rightarrow Bphe_A$. If  available and neutral, $Q_B$ is reduced, but if it was already reduced, the anionic species $Q_A^-Q_B^-$ become a two-proton acceptor that results in a Quinol $Q_BH_2$. The Quinol dissociates from the RC in order to deliver protons to the intracytoplasm and increase the voltage that drives the ATP synthesis \cite{Graige,Osvath}, driven concomitantly by the oxidation of the cytochrome  $cyt^2\rightarrow cyt^3$ that restablishes  the neutrality of $P$ after photo-excitation.  The participation of the main reactants of quinone and cytochrome cycles in the charge transfer dynamics along the active branch is shown in the lower part of (b).}
		\label{fig1}
\end{figure}

\begin{figure}
		\centering
			\includegraphics[scale=0.9]{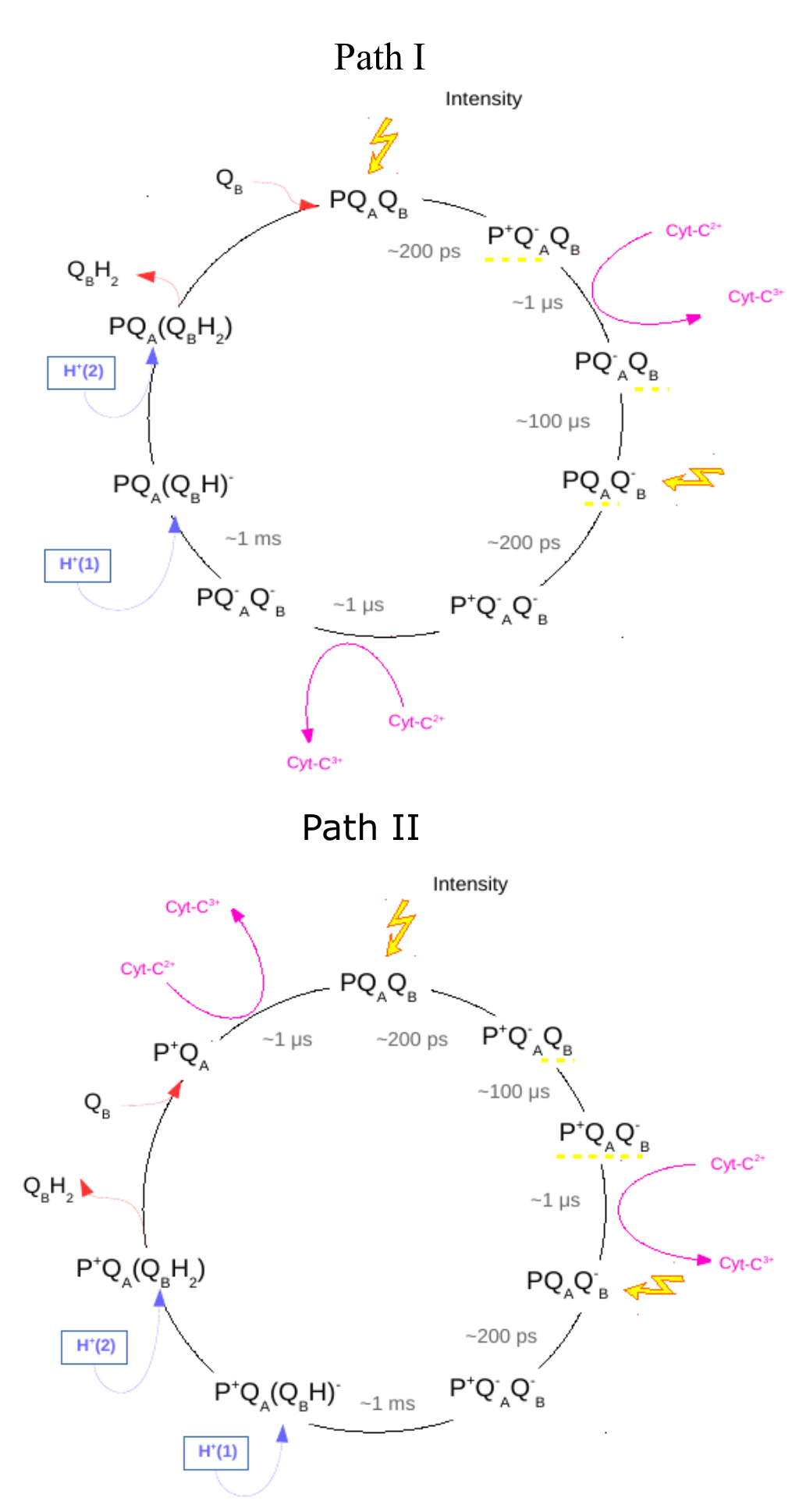}
		\caption{\footnotesize{Steps and time scales followed in paths II and II shown in Figure \ref{fig1}(b).}}
		\label{paths}
	\end{figure}

Photosynthesis captures and processes thermal light in order to produce ATP.  Initial  light harvesting is followed by subsequent excitonic transfer to RCs, where the electronic excitation of the special pair (P) chromophores within the RC, $P\rightarrow P^*$, induces a charge separation of the special pair $P^*\rightarrow P^++e$. This electron $e$ is transferred along an active branch  of the RC as shown in Fig.\ref{fig1}(a), reducing several cofactors until it reduces a quinone molecule $Q_A\rightarrow Q_A^-$. If an auxiliary quinone $Q_B$ is available,  the  reduction  $Q_A^-Q_B\rightarrow Q_AQ_B^-$ takes place. The oxidised $P^+$, after becoming  neutral, can reduce the quinone $Q_A$ again, $P^*Q_AQ_B^-\rightarrow P^+Q_A^-Q_B^-$,  in order to proceed with the uptake of two intracytoplasmic protons $Q_A^-Q_B^-+2 H^+\rightarrow Q_A(Q_BH_2)$, i.e., two $Q_A$ reductions take place before a quinol molecule is synthesized. Each reduction has two possible pathways I and II depending on the availability of quinones and cytochromes, as displayed in \ref{fig1}(b)-(c). All the steps and time scales involved are presented in more detail in Fig.\ref{paths}. In a later stage, with the help of the cytochrome complex, quinol $Q_BH_2$ releases these protons and other two protons are pumped from the outside of the vesicle. The electrochemical gradient produced by these four protons is subsequently used by a transmembrane macromolecule called ATP-synthase (ATPase) to produce an ATP molecule. Intermediate metastable states occur between the first $Q_A^-Q_B\rightarrow Q_AQ_B^-$ and second $ Q_A^-Q_B^-$ reduction of the quinone pair. In particular, the metastable states $P^+Q_A^-$  or $P^+Q_B^-$ will degrade to $PQ_A$ or $PQ_B$ due to charge recombination in about 100 ms or 1000 ms, respectively \cite{Graige,Diner, Milano}. The $P^+Q_B^-$  recombination is of special relevance, since the dynamics within the RCs will more often populate this state between consecutive $P$ photo-excitations \cite{Graige}.  While the absorption of photon pairs may help to avoid the decay of the metastable state, quinol production may be inhibited whenever excitations that reach a given RC are delayed longer than charge recombination.

\section{Effect of the spatial correlations}\label{spc}

We use equation \ref{Fdt} to get the detection events in the full set of LHs, and propose a  probability density function (PDF) to place subsequent photo-excitations $P(\nu_{i,j})$, where $\nu_{i,j}=2\pi\vert \vec{r}_i-\vec{r}_j \vert /l_c$ compares the distance between detectors to the coherence length (for simplicity we will write $\nu$ from now on). This probability is defined as the normalized second order spatial coherence of the intensity

\begin{equation}
P(\nu)=\frac{g^{(2)}(\nu)}{\int_0^D d\nu g^{(2)}(\nu)} \text{ ,}
\end{equation}

\noindent with $\left( g^{(2)}(\nu)=1+\vert 2\frac{J_1(\nu)}{\nu} \vert^{2}\right)$ \cite{Mandel}, and $D$ is the diameter of the photosynthetic membrane. Since each type of light harvesting complex absorbs at different frequency, the light absorption is simulated independently for the LH1 and LH2 networks. Then, the two time series are combined to get the absorption signal that enters into the next stages of the photosynthesis simulations. In our numerical implementation, after a photo-excitation is located at the position $\vec{r}_i$ of an LH1 or LH2 complex, the next photo-excitation that goes to the same LH type is placed based upon $P(\nu)$. The position for the next photo-excitation $\vec{r}_l$ is selected to lie in a radius $\vert \vec{r}_k-\vec{r}_l\vert$, closest to $\nu$.\\

\begin{figure}
		\centering
		\vspace{0.5cm}
   		\includegraphics[width=1\columnwidth]{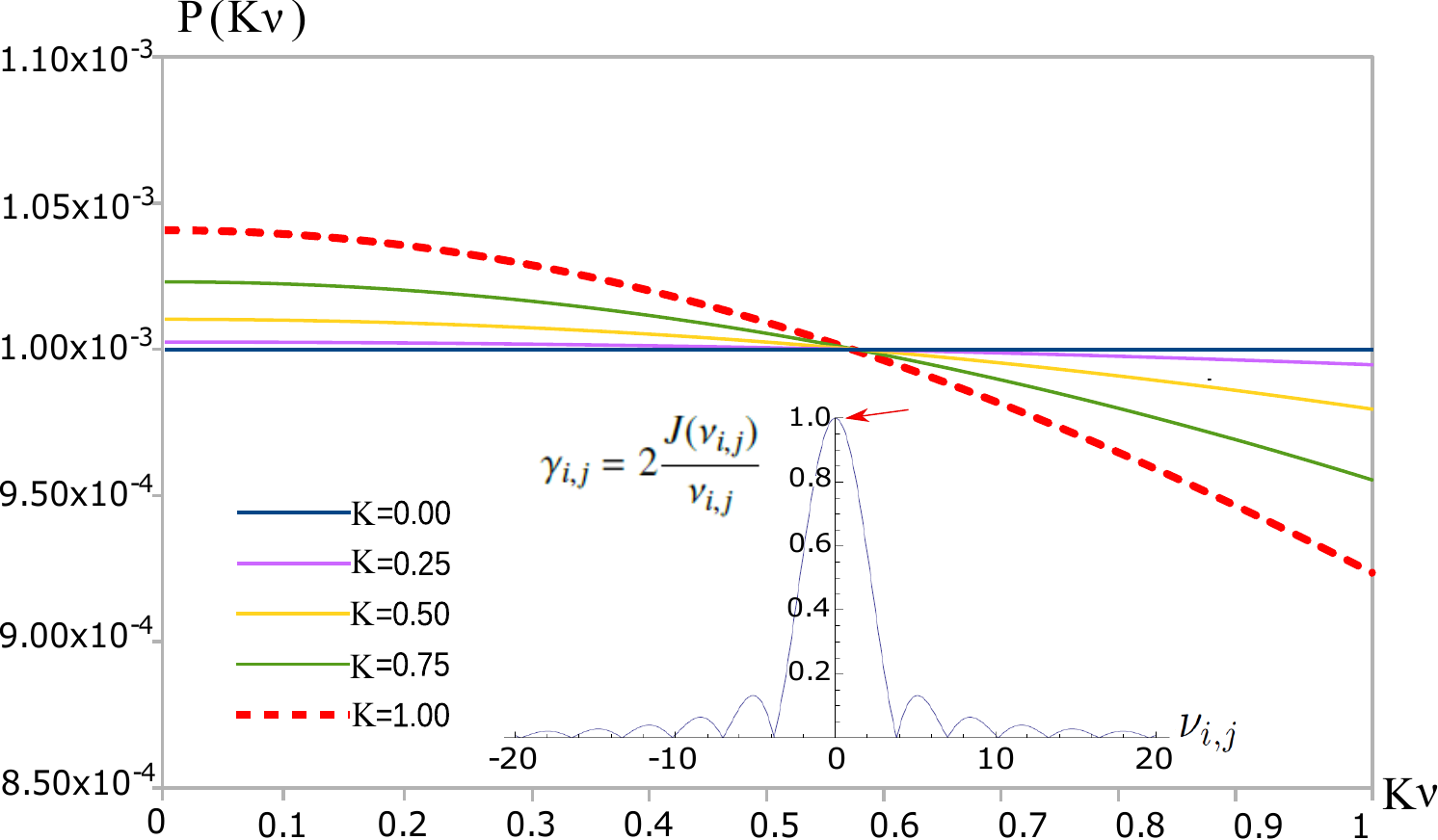}
		\caption{\footnotesize{Spatial probability distribution $P(F \nu)$, for different $K$ values. Inset: Spatial coherence function $\gamma_{i,j}(\nu_{i,j})$. The red arrow points the spot in the spacial coherence function where the probability distribution tends to be constant.}}
		\label{Sp}
	\end{figure}

The spatial correlations degree is tested by scaling the radius within which the next photon is located ($\nu \rightarrow K\nu$). The PDF plotted in fig.\ref{Sp} shows a profile that flattens out as $K \rightarrow 0$ to a uniform distribution. This happens because the detection area gets much smaller than the coherence area ($\nu \ll 1 \; \forall i,j$) and therefore the spatial coherence becomes very close to its maximum value where the slope of the function goes to zero (see inset in fig.\ref{Sp}). Particularly, for the actual size of the membrane ($K\nu \leq 0.01$), $P(\nu)$ is practically constant throughout the full extent of the membrane because of the high degree of spatial correlations. This means, that inside of such a small detection area the arrivals are randomly distributed, but the photocounting statistics are not Poissonian. Note that only when the spatial and temporal correlations are present ($l \lesssim lc$ and $T \lesssim \tau_c$), the bunched structure of thermal light manifests (see fig.2(a) in the main text).

	 \begin{figure}[]
		\centering
			\includegraphics[width=\columnwidth]{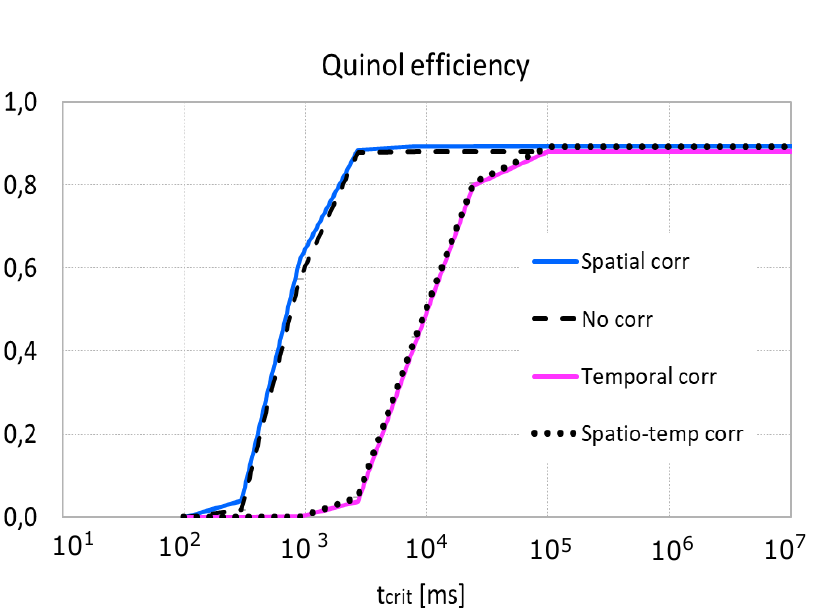}
		\caption{\footnotesize{Quinol production efficiency for 4 cases: (Dashed) no correlations at all, (Magenta) The incoming signal presents burstiness but the photons are randomly distributed over the membrane, (Blue) Only spatial correlations but the incoming signal is Poissonian, (Dots) Present spatial and temporal correlations included in simulations. Horizontal axis in logarithmic scale. In these simulations $ \langle I\rangle=10^{-3}$photons/(LH$\cdot$ms), and $10^{6}$ light absorption events.}}
		\label{Tcrit}
	\end{figure}

To work out the effects of spatial and temporal correlations separately, we first calculated the proportion of exciton-pairs converted into quinol, for the 4 cases depicted in Fig.\ref{Tcrit}: 1) No correlations at all (Poissonian incoming signal). 2) Only temporal correlations with random spatial delivering. 3) Only spatially correlated placement of excitations, without temporal burstiness. 4) Spatio-temporal correlations occurring together as it actually is. In all the plots the \textit{non-correlation} and \textit{only-spatial correlation} curves overlap, in the same way as \textit{temporal} and \textit{spatio-temporal correlation} curves do. These results point out how the temporal correlations are responsible for the changes in quinol production, since the spatial correlations are very high and almost constant through the membrane. These results corroborate that the spatial assignment of the arrivals does not affect reception statistics (and therefore the quinol efficiency), because the entire vesicle is very small compared with the light's coherence area (the vesicle's area is 10000 smaller than the coherence area), thus behaving like a punctual receptor. Therefore, the internal configuration of detectors is irrelevant for the absorption statistics and  for the quinol production.


\begin{thebibliography}{64}%
\makeatletter
\providecommand \@ifxundefined [1]{%
 \@ifx{#1\undefined}
}%
\providecommand \@ifnum [1]{%
 \ifnum #1\expandafter \@firstoftwo
 \else \expandafter \@secondoftwo
 \fi
}%
\providecommand \@ifx [1]{%
 \ifx #1\expandafter \@firstoftwo
 \else \expandafter \@secondoftwo
 \fi
}%
\providecommand \natexlab [1]{#1}%
\providecommand \enquote  [1]{``#1''}%
\providecommand \bibnamefont  [1]{#1}%
\providecommand \bibfnamefont [1]{#1}%
\providecommand \citenamefont [1]{#1}%
\providecommand \href@noop [0]{\@secondoftwo}%
\providecommand \href [0]{\begingroup \@sanitize@url \@href}%
\providecommand \@href[1]{\@@startlink{#1}\@@href}%
\providecommand \@@href[1]{\endgroup#1\@@endlink}%
\providecommand \@sanitize@url [0]{\catcode `\\12\catcode `\$12\catcode
  `\&12\catcode `\#12\catcode `\^12\catcode `\_12\catcode `\%12\relax}%
\providecommand \@@startlink[1]{}%
\providecommand \@@endlink[0]{}%
\providecommand \url  [0]{\begingroup\@sanitize@url \@url }%
\providecommand \@url [1]{\endgroup\@href {#1}{\urlprefix }}%
\providecommand \urlprefix  [0]{URL }%
\providecommand \Eprint [0]{\href }%
\providecommand \doibase [0]{http://dx.doi.org/}%
\providecommand \selectlanguage [0]{\@gobble}%
\providecommand \bibinfo  [0]{\@secondoftwo}%
\providecommand \bibfield  [0]{\@secondoftwo}%
\providecommand \translation [1]{[#1]}%
\providecommand \BibitemOpen [0]{}%
\providecommand \bibitemStop [0]{}%
\providecommand \bibitemNoStop [0]{.\EOS\space}%
\providecommand \EOS [0]{\spacefactor3000\relax}%
\providecommand \BibitemShut  [1]{\csname bibitem#1\endcsname}%
\let\auto@bib@innerbib\@empty
\bibitem [{\citenamefont {Scheuring}\ and\ \citenamefont
  {Sturgis}(2005)}]{Scheuring}%
  \BibitemOpen
  \bibfield  {author} {\bibinfo {author} {\bibfnamefont {S.}~\bibnamefont
  {Scheuring}}\ and\ \bibinfo {author} {\bibfnamefont {J.~N.}\ \bibnamefont
  {Sturgis}},\ }\href {\doibase 10.1126/science.1110879} {\bibfield  {journal}
  {\bibinfo  {journal} {Science}\ }\textbf {\bibinfo {volume} {309}},\ \bibinfo
  {pages} {484} (\bibinfo {year} {2005})}\BibitemShut {NoStop}%
\bibitem [{\citenamefont {Geyer}\ and\ \citenamefont {Helms}(2006)}]{Geyer}%
  \BibitemOpen
  \bibfield  {author} {\bibinfo {author} {\bibfnamefont {T.}~\bibnamefont
  {Geyer}}\ and\ \bibinfo {author} {\bibfnamefont {V.}~\bibnamefont {Helms}},\
  }\href {\doibase 10.1529/biophysj.105.078501} {\bibfield  {journal} {\bibinfo
   {journal} {Biophysical Journal}\ }\textbf {\bibinfo {volume} {91}},\
  \bibinfo {pages} {921} (\bibinfo {year} {2006})}\BibitemShut {NoStop}%
\bibitem [{\citenamefont {Cogdell}\ \emph {et~al.}(2000)\citenamefont
  {Cogdell}, \citenamefont {Howard}, \citenamefont {Bittl}, \citenamefont
  {Schlodder}, \citenamefont {Geisenheimer},\ and\ \citenamefont
  {Lubitz}}]{Cogdell1}%
  \BibitemOpen
  \bibfield  {author} {\bibinfo {author} {\bibfnamefont {R.~J.}\ \bibnamefont
  {Cogdell}}, \bibinfo {author} {\bibfnamefont {T.~D.}\ \bibnamefont {Howard}},
  \bibinfo {author} {\bibfnamefont {R.}~\bibnamefont {Bittl}}, \bibinfo
  {author} {\bibfnamefont {E.}~\bibnamefont {Schlodder}}, \bibinfo {author}
  {\bibfnamefont {I.}~\bibnamefont {Geisenheimer}}, \ and\ \bibinfo {author}
  {\bibfnamefont {W.}~\bibnamefont {Lubitz}},\ }\href {\doibase
  10.1098/rstb.2000.0696} {\bibfield  {journal} {\bibinfo  {journal}
  {Philosophical Transactions of the Royal Society B: Biological Sciences}\
  }\textbf {\bibinfo {volume} {355}},\ \bibinfo {pages} {1345} (\bibinfo {year}
  {2000})}\BibitemShut {NoStop}%
\bibitem [{\citenamefont {Kim}\ \emph {et~al.}(2007)\citenamefont {Kim},
  \citenamefont {Li}, \citenamefont {Maresca}, \citenamefont {Bryant},\ and\
  \citenamefont {Savikhin}}]{Kim}%
  \BibitemOpen
  \bibfield  {author} {\bibinfo {author} {\bibfnamefont {H.}~\bibnamefont
  {Kim}}, \bibinfo {author} {\bibfnamefont {H.}~\bibnamefont {Li}}, \bibinfo
  {author} {\bibfnamefont {J.~A.}\ \bibnamefont {Maresca}}, \bibinfo {author}
  {\bibfnamefont {D.~A.}\ \bibnamefont {Bryant}}, \ and\ \bibinfo {author}
  {\bibfnamefont {S.}~\bibnamefont {Savikhin}},\ }\href {\doibase
  10.1529/biophysj.106.103556} {\bibfield  {journal} {\bibinfo  {journal}
  {Biophysical Journal}\ }\textbf {\bibinfo {volume} {93}},\ \bibinfo {pages}
  {192} (\bibinfo {year} {2007})}\BibitemShut {NoStop}%
\bibitem [{\citenamefont {Scholes}\ \emph {et~al.}(1997)\citenamefont
  {Scholes}, \citenamefont {Harcourt},\ and\ \citenamefont
  {Fleming}}]{Fleming1}%
  \BibitemOpen
  \bibfield  {author} {\bibinfo {author} {\bibfnamefont {G.~D.}\ \bibnamefont
  {Scholes}}, \bibinfo {author} {\bibfnamefont {R.~D.}\ \bibnamefont
  {Harcourt}}, \ and\ \bibinfo {author} {\bibfnamefont {G.~R.}\ \bibnamefont
  {Fleming}},\ }\href {\doibase 10.1021/jp963970a} {\bibfield  {journal}
  {\bibinfo  {journal} {The Journal of Physical Chemistry B}\ }\textbf
  {\bibinfo {volume} {101}},\ \bibinfo {pages} {7302} (\bibinfo {year}
  {1997})}\BibitemShut {NoStop}%
\bibitem [{\citenamefont {Kosumi}\ \emph {et~al.}(2016)\citenamefont {Kosumi},
  \citenamefont {Horibe}, \citenamefont {Sugisaki}, \citenamefont {Cogdell},\
  and\ \citenamefont {Hashimoto}}]{Kosumi}%
  \BibitemOpen
  \bibfield  {author} {\bibinfo {author} {\bibfnamefont {D.}~\bibnamefont
  {Kosumi}}, \bibinfo {author} {\bibfnamefont {T.}~\bibnamefont {Horibe}},
  \bibinfo {author} {\bibfnamefont {M.}~\bibnamefont {Sugisaki}}, \bibinfo
  {author} {\bibfnamefont {R.~J.}\ \bibnamefont {Cogdell}}, \ and\ \bibinfo
  {author} {\bibfnamefont {H.}~\bibnamefont {Hashimoto}},\ }\href {\doibase
  10.1021/acs.jpcb.6b00121} {\bibfield  {journal} {\bibinfo  {journal} {The
  Journal of Physical Chemistry B}\ }\textbf {\bibinfo {volume} {120}},\
  \bibinfo {pages} {951} (\bibinfo {year} {2016})},\ \bibinfo {note} {pMID:
  26800035}\BibitemShut {NoStop}%
\bibitem [{\citenamefont {Dong}\ \emph {et~al.}(2017)\citenamefont {Dong},
  \citenamefont {Li}, \citenamefont {Yi}, \citenamefont {Agarwal},\ and\
  \citenamefont {Scully}}]{Scully-blockade}%
  \BibitemOpen
  \bibfield  {author} {\bibinfo {author} {\bibfnamefont {H.}~\bibnamefont
  {Dong}}, \bibinfo {author} {\bibfnamefont {S.-W.}\ \bibnamefont {Li}},
  \bibinfo {author} {\bibfnamefont {Z.}~\bibnamefont {Yi}}, \bibinfo {author}
  {\bibfnamefont {G.~S.}\ \bibnamefont {Agarwal}}, \ and\ \bibinfo {author}
  {\bibfnamefont {M.~O.}\ \bibnamefont {Scully}},\ }\href
  {https://arxiv.org/pdf/1608.04364.pdf} {\bibfield  {journal} {\bibinfo
  {journal} {arXiv:1608.04364v2 [physics.chem-ph]}\ } (\bibinfo {year}
  {2017})}\BibitemShut {NoStop}%
\bibitem [{\citenamefont {Bures}\ \emph
  {et~al.}(1972{\natexlab{a}})\citenamefont {Bures}, \citenamefont {Delisle},\
  and\ \citenamefont {Zardecki}}]{Bures71}%
  \BibitemOpen
  \bibfield  {author} {\bibinfo {author} {\bibfnamefont {J.}~\bibnamefont
  {Bures}}, \bibinfo {author} {\bibfnamefont {C.}~\bibnamefont {Delisle}}, \
  and\ \bibinfo {author} {\bibfnamefont {A.}~\bibnamefont {Zardecki}},\ }\href
  {\doibase 10.1139/p72-108} {\bibfield  {journal} {\bibinfo  {journal} {Can.
  J. Phys.}\ }\textbf {\bibinfo {volume} {50}},\ \bibinfo {pages} {760}
  (\bibinfo {year} {1972}{\natexlab{a}})}\BibitemShut {NoStop}%
\bibitem [{\citenamefont {Mashaal}\ \emph {et~al.}(2012)\citenamefont
  {Mashaal}, \citenamefont {Goldstein}, \citenamefont {Feuermann},\ and\
  \citenamefont {Gordon}}]{Mashaal2}%
  \BibitemOpen
  \bibfield  {author} {\bibinfo {author} {\bibfnamefont {H.}~\bibnamefont
  {Mashaal}}, \bibinfo {author} {\bibfnamefont {A.}~\bibnamefont {Goldstein}},
  \bibinfo {author} {\bibfnamefont {D.}~\bibnamefont {Feuermann}}, \ and\
  \bibinfo {author} {\bibfnamefont {J.~M.}\ \bibnamefont {Gordon}},\ }\href
  {\doibase 10.1364/ol.37.003516} {\bibfield  {journal} {\bibinfo  {journal}
  {Opt. Lett.}\ }\textbf {\bibinfo {volume} {37}},\ \bibinfo {pages} {3516}
  (\bibinfo {year} {2012})}\BibitemShut {NoStop}%
\bibitem [{\citenamefont {Kano}\ and\ \citenamefont {Wolf}(1962)}]{Kano}%
  \BibitemOpen
  \bibfield  {author} {\bibinfo {author} {\bibfnamefont {Y.}~\bibnamefont
  {Kano}}\ and\ \bibinfo {author} {\bibfnamefont {E.}~\bibnamefont {Wolf}},\
  }\href {\doibase 10.1088/0370-1328/80/6/308} {\bibfield  {journal} {\bibinfo
  {journal} {Proceedings of the Physical Society}\ }\textbf {\bibinfo {volume}
  {80}},\ \bibinfo {pages} {1273} (\bibinfo {year} {1962})}\BibitemShut
  {NoStop}%
\bibitem [{\citenamefont {Glauber}(1963)}]{Glauber_1963PR}%
  \BibitemOpen
  \bibfield  {author} {\bibinfo {author} {\bibfnamefont {R.~J.}\ \bibnamefont
  {Glauber}},\ }\href {\doibase 10.1103/PhysRev.130.2529} {\bibfield  {journal}
  {\bibinfo  {journal} {Phys. Rev.}\ }\textbf {\bibinfo {volume} {130}},\
  \bibinfo {pages} {2529} (\bibinfo {year} {1963})}\BibitemShut {NoStop}%
\bibitem [{\citenamefont {Mandel}\ \emph {et~al.}(1964)\citenamefont {Mandel},
  \citenamefont {Sudarshan},\ and\ \citenamefont {Wolf}}]{Mandel_1964PPS}%
  \BibitemOpen
  \bibfield  {author} {\bibinfo {author} {\bibfnamefont {L.}~\bibnamefont
  {Mandel}}, \bibinfo {author} {\bibfnamefont {E.~C.~G.}\ \bibnamefont
  {Sudarshan}}, \ and\ \bibinfo {author} {\bibfnamefont {E.}~\bibnamefont
  {Wolf}},\ }\href {\doibase 10.1088/0370-1328/84/3/313} {\bibfield  {journal}
  {\bibinfo  {journal} {Proceedings of the Physical Society}\ }\textbf
  {\bibinfo {volume} {84}},\ \bibinfo {pages} {435} (\bibinfo {year}
  {1964})}\BibitemShut {NoStop}%
\bibitem [{\citenamefont {Jang}\ and\ \citenamefont
  {Silbey}(2003)}]{Silbey_JCP2003}%
  \BibitemOpen
  \bibfield  {author} {\bibinfo {author} {\bibfnamefont {S.}~\bibnamefont
  {Jang}}\ and\ \bibinfo {author} {\bibfnamefont {R.~J.}\ \bibnamefont
  {Silbey}},\ }\href {\doibase 10.1063/1.1569239} {\bibfield  {journal}
  {\bibinfo  {journal} {The Journal of Chemical Physics}\ }\textbf {\bibinfo
  {volume} {118}},\ \bibinfo {pages} {9312} (\bibinfo {year}
  {2003})}\BibitemShut {NoStop}%
\bibitem [{\citenamefont {{Bergstrom}}\ \emph {et~al.}(1989)\citenamefont
  {{Bergstrom}}, \citenamefont {{van Grondelle}},\ and\ \citenamefont
  {{Sunsdst\"om}}}]{Bergstrom_1989FEBS}%
  \BibitemOpen
  \bibfield  {author} {\bibinfo {author} {\bibfnamefont {H.}~\bibnamefont
  {{Bergstrom}}}, \bibinfo {author} {\bibfnamefont {R.}~\bibnamefont {{van
  Grondelle}}}, \ and\ \bibinfo {author} {\bibfnamefont {V.}~\bibnamefont
  {{Sunsdst\"om}}},\ }\href {https://doi.org/10.1016/0014-5793(89)80785-9}
  {\bibfield  {journal} {\bibinfo  {journal} {FEBS Lett.}\ }\textbf {\bibinfo
  {volume} {250}},\ \bibinfo {pages} {503} (\bibinfo {year}
  {1989})}\BibitemShut {NoStop}%
\bibitem [{\citenamefont {Timpmann}\ \emph {et~al.}(1993)\citenamefont
  {Timpmann}, \citenamefont {Zhang}, \citenamefont {Freiberg},\ and\
  \citenamefont {Sundstr{\"o}m}}]{Timpmann}%
  \BibitemOpen
  \bibfield  {author} {\bibinfo {author} {\bibfnamefont {K.}~\bibnamefont
  {Timpmann}}, \bibinfo {author} {\bibfnamefont {F.~G.}\ \bibnamefont {Zhang}},
  \bibinfo {author} {\bibfnamefont {A.}~\bibnamefont {Freiberg}}, \ and\
  \bibinfo {author} {\bibfnamefont {V.}~\bibnamefont {Sundstr{\"o}m}},\ }\href
  {\doibase 10.1016/0005-2728(93)90017-a} {\bibfield  {journal} {\bibinfo
  {journal} {Biochimica et Biophysica Acta}\ }\textbf {\bibinfo {volume}
  {1183}},\ \bibinfo {pages} {185} (\bibinfo {year} {1993})}\BibitemShut
  {NoStop}%
\bibitem [{\citenamefont {van Grondelle}\ \emph {et~al.}(1994)\citenamefont
  {van Grondelle}, \citenamefont {Dekker}, \citenamefont {Gillbro},\ and\
  \citenamefont {Sundstr{\"o}m}}]{Vgrondelle}%
  \BibitemOpen
  \bibfield  {author} {\bibinfo {author} {\bibfnamefont {R.}~\bibnamefont {van
  Grondelle}}, \bibinfo {author} {\bibfnamefont {J.~P.}\ \bibnamefont
  {Dekker}}, \bibinfo {author} {\bibfnamefont {T.}~\bibnamefont {Gillbro}}, \
  and\ \bibinfo {author} {\bibfnamefont {V.}~\bibnamefont {Sundstr{\"o}m}},\
  }\href {\doibase 10.1016/0005-2728(94)90166-x} {\bibfield  {journal}
  {\bibinfo  {journal} {Biochimica et Biophysica Acta}\ }\textbf {\bibinfo
  {volume} {1187}},\ \bibinfo {pages} {1} (\bibinfo {year} {1994})}\BibitemShut
  {NoStop}%
\bibitem [{\citenamefont {{Visscher}}\ \emph {et~al.}(1989)\citenamefont
  {{Visscher}}, \citenamefont {{Bergstr{\"o}m}}, \citenamefont
  {{S{\"u}ndstrom}}, \citenamefont {{Hunter}},\ and\ \citenamefont {{van
  Grondelle}}}]{Visscher_1989PhotRes}%
  \BibitemOpen
  \bibfield  {author} {\bibinfo {author} {\bibfnamefont {K.}~\bibnamefont
  {{Visscher}}}, \bibinfo {author} {\bibfnamefont {H.}~\bibnamefont
  {{Bergstr{\"o}m}}}, \bibinfo {author} {\bibfnamefont {V.}~\bibnamefont
  {{S{\"u}ndstrom}}}, \bibinfo {author} {\bibfnamefont {C.}~\bibnamefont
  {{Hunter}}}, \ and\ \bibinfo {author} {\bibfnamefont {R.}~\bibnamefont {{van
  Grondelle}}},\ }\href {https://doi.org/10.1007/BF00048300} {\bibfield
  {journal} {\bibinfo  {journal} {Photosynthesys Research}\ }\textbf {\bibinfo
  {volume} {22}},\ \bibinfo {pages} {211} (\bibinfo {year} {1989})}\BibitemShut
  {NoStop}%
\bibitem [{\citenamefont {{Harel}}\ and\ \citenamefont
  {{Engel}}(2012)}]{Engel_PNAS2012}%
  \BibitemOpen
  \bibfield  {author} {\bibinfo {author} {\bibfnamefont {E.}~\bibnamefont
  {{Harel}}}\ and\ \bibinfo {author} {\bibfnamefont {G.~S.}\ \bibnamefont
  {{Engel}}},\ }\href {https://www.pnas.org/content/109/3/706.short} {\bibfield
   {journal} {\bibinfo  {journal} {Proc. Natl Acad. Sci. USA}\ }\textbf
  {\bibinfo {volume} {109}},\ \bibinfo {pages} {706} (\bibinfo {year}
  {2012})}\BibitemShut {NoStop}%
\bibitem [{\citenamefont {{Fuller}}\ \emph {et~al.}(2014)\citenamefont
  {{Fuller}}, \citenamefont {{Pan}}, \citenamefont {{Gelzinis}}, \citenamefont
  {{Butkus}}, \citenamefont {{Senlik}}, \citenamefont {{Wilcox}}, \citenamefont
  {{Yocum}}, \citenamefont {{Valkunas}}, \citenamefont {{Abramavicius}},\ and\
  \citenamefont {{Ogilvie}}}]{Ogilvie_NChem2014}%
  \BibitemOpen
  \bibfield  {author} {\bibinfo {author} {\bibfnamefont {F.~D.}\ \bibnamefont
  {{Fuller}}}, \bibinfo {author} {\bibfnamefont {J.}~\bibnamefont {{Pan}}},
  \bibinfo {author} {\bibfnamefont {A.}~\bibnamefont {{Gelzinis}}}, \bibinfo
  {author} {\bibfnamefont {V.}~\bibnamefont {{Butkus}}}, \bibinfo {author}
  {\bibfnamefont {S.~S.}\ \bibnamefont {{Senlik}}}, \bibinfo {author}
  {\bibfnamefont {D.~E.}\ \bibnamefont {{Wilcox}}}, \bibinfo {author}
  {\bibfnamefont {C.~F.}\ \bibnamefont {{Yocum}}}, \bibinfo {author}
  {\bibfnamefont {L.}~\bibnamefont {{Valkunas}}}, \bibinfo {author}
  {\bibfnamefont {D.}~\bibnamefont {{Abramavicius}}}, \ and\ \bibinfo {author}
  {\bibfnamefont {J.~P.}\ \bibnamefont {{Ogilvie}}},\ }\href
  {https://doi.org/10.1038/nchem.2005} {\bibfield  {journal} {\bibinfo
  {journal} {Nature Chem.}\ }\textbf {\bibinfo {volume} {6}},\ \bibinfo {pages}
  {706} (\bibinfo {year} {2014})}\BibitemShut {NoStop}%
\bibitem [{\citenamefont {Thyrhaug}\ \emph {et~al.}(2018)\citenamefont
  {Thyrhaug}, \citenamefont {Tempelaar}, \citenamefont {Alcocer}, \citenamefont
  {{\v Z}{\'\i}dek}, \citenamefont {B{\'\i}na}, \citenamefont {Knoester},
  \citenamefont {Jansen},\ and\ \citenamefont
  {Zigmantas}}]{Zigmantas_2018NatChem}%
  \BibitemOpen
  \bibfield  {author} {\bibinfo {author} {\bibfnamefont {E.}~\bibnamefont
  {Thyrhaug}}, \bibinfo {author} {\bibfnamefont {R.}~\bibnamefont {Tempelaar}},
  \bibinfo {author} {\bibfnamefont {M.~J.~P.}\ \bibnamefont {Alcocer}},
  \bibinfo {author} {\bibfnamefont {K.}~\bibnamefont {{\v Z}{\'\i}dek}},
  \bibinfo {author} {\bibfnamefont {D.}~\bibnamefont {B{\'\i}na}}, \bibinfo
  {author} {\bibfnamefont {J.}~\bibnamefont {Knoester}}, \bibinfo {author}
  {\bibfnamefont {T.~L.~C.}\ \bibnamefont {Jansen}}, \ and\ \bibinfo {author}
  {\bibfnamefont {D.}~\bibnamefont {Zigmantas}},\ }\href {\doibase
  10.1038/s41557-018-0060-5} {\bibfield  {journal} {\bibinfo  {journal} {Nature
  Chemistry}\ }\textbf {\bibinfo {volume} {10}},\ \bibinfo {pages} {780}
  (\bibinfo {year} {2018})}\BibitemShut {NoStop}%
\bibitem [{\citenamefont {Romero}\ \emph {et~al.}(2014)\citenamefont {Romero},
  \citenamefont {Augulis}, \citenamefont {Novoderezhkin}, \citenamefont
  {Ferretti}, \citenamefont {Thieme}, \citenamefont {Zigmantas},\ and\
  \citenamefont {van Grondelle}}]{vanGrondelle2}%
  \BibitemOpen
  \bibfield  {author} {\bibinfo {author} {\bibfnamefont {E.}~\bibnamefont
  {Romero}}, \bibinfo {author} {\bibfnamefont {R.}~\bibnamefont {Augulis}},
  \bibinfo {author} {\bibfnamefont {V.~I.}\ \bibnamefont {Novoderezhkin}},
  \bibinfo {author} {\bibfnamefont {M.}~\bibnamefont {Ferretti}}, \bibinfo
  {author} {\bibfnamefont {J.}~\bibnamefont {Thieme}}, \bibinfo {author}
  {\bibfnamefont {D.}~\bibnamefont {Zigmantas}}, \ and\ \bibinfo {author}
  {\bibfnamefont {R.}~\bibnamefont {van Grondelle}},\ }\href
  {https://www.nature.com/articles/nphys3017} {\bibfield  {journal} {\bibinfo
  {journal} {Nature Physics}\ }\textbf {\bibinfo {volume} {10}},\ \bibinfo
  {pages} {676} (\bibinfo {year} {2014})}\BibitemShut {NoStop}%
\bibitem [{\citenamefont {Graige}\ \emph {et~al.}(1996)\citenamefont {Graige},
  \citenamefont {Paddock}, \citenamefont {Bruce}, \citenamefont {Feher},\ and\
  \citenamefont {Okamura}}]{Graige}%
  \BibitemOpen
  \bibfield  {author} {\bibinfo {author} {\bibfnamefont {M.~S.}\ \bibnamefont
  {Graige}}, \bibinfo {author} {\bibfnamefont {M.~L.}\ \bibnamefont {Paddock}},
  \bibinfo {author} {\bibfnamefont {J.~M.}\ \bibnamefont {Bruce}}, \bibinfo
  {author} {\bibfnamefont {G.}~\bibnamefont {Feher}}, \ and\ \bibinfo {author}
  {\bibfnamefont {M.~Y.}\ \bibnamefont {Okamura}},\ }\href {\doibase
  10.1021/ja960056m} {\bibfield  {journal} {\bibinfo  {journal} {Journal of the
  American Chemical Society}\ }\textbf {\bibinfo {volume} {118}},\ \bibinfo
  {pages} {9005} (\bibinfo {year} {1996})}\BibitemShut {NoStop}%
\bibitem [{\citenamefont {Diner}\ \emph {et~al.}(1984)\citenamefont {Diner},
  \citenamefont {Schenck},\ and\ \citenamefont {Vitry}}]{Diner}%
  \BibitemOpen
  \bibfield  {author} {\bibinfo {author} {\bibfnamefont {B.~A.}\ \bibnamefont
  {Diner}}, \bibinfo {author} {\bibfnamefont {C.~C.}\ \bibnamefont {Schenck}},
  \ and\ \bibinfo {author} {\bibfnamefont {D.}~\bibnamefont {Vitry}},\ }\href
  {https://doi.org/10.1016/0005-2728(84)90211-1} {\bibfield  {journal}
  {\bibinfo  {journal} {Biochimica et Biophysica Acta}\ }\textbf {\bibinfo
  {volume} {766}},\ \bibinfo {pages} {9} (\bibinfo {year} {1984})}\BibitemShut
  {NoStop}%
\bibitem [{\citenamefont {Milano}\ \emph {et~al.}(2003)\citenamefont {Milano},
  \citenamefont {Agostiano}, \citenamefont {Mavelli},\ and\ \citenamefont
  {Trotta}}]{Milano}%
  \BibitemOpen
  \bibfield  {author} {\bibinfo {author} {\bibfnamefont {F.}~\bibnamefont
  {Milano}}, \bibinfo {author} {\bibfnamefont {A.}~\bibnamefont {Agostiano}},
  \bibinfo {author} {\bibfnamefont {F.}~\bibnamefont {Mavelli}}, \ and\
  \bibinfo {author} {\bibfnamefont {M.}~\bibnamefont {Trotta}},\ }\href
  {\doibase 10.1046/j.1432-1033.2003.03845.x} {\bibfield  {journal} {\bibinfo
  {journal} {European Journal of Biochemistry}\ }\textbf {\bibinfo {volume}
  {270}},\ \bibinfo {pages} {4595} (\bibinfo {year} {2003})}\BibitemShut
  {NoStop}%
\bibitem [{\citenamefont {Frese}\ \emph {et~al.}(2004)\citenamefont {Frese},
  \citenamefont {Siebert}, \citenamefont {Niederman}, \citenamefont {Hunter},
  \citenamefont {Otto},\ and\ \citenamefont {van
  Grondelle}}]{vanGrondelle_PNAS2004}%
  \BibitemOpen
  \bibfield  {author} {\bibinfo {author} {\bibfnamefont {R.~N.}\ \bibnamefont
  {Frese}}, \bibinfo {author} {\bibfnamefont {C.~A.}\ \bibnamefont {Siebert}},
  \bibinfo {author} {\bibfnamefont {R.~A.}\ \bibnamefont {Niederman}}, \bibinfo
  {author} {\bibfnamefont {C.~N.}\ \bibnamefont {Hunter}}, \bibinfo {author}
  {\bibfnamefont {C.}~\bibnamefont {Otto}}, \ and\ \bibinfo {author}
  {\bibfnamefont {R.}~\bibnamefont {van Grondelle}},\ }\href {\doibase
  10.1073/pnas.0407295102} {\bibfield  {journal} {\bibinfo  {journal}
  {Proceedings of the National Academy of Sciences}\ }\textbf {\bibinfo
  {volume} {101}},\ \bibinfo {pages} {17994} (\bibinfo {year}
  {2004})}\BibitemShut {NoStop}%
\bibitem [{\citenamefont {Caycedo-Soler}\ \emph {et~al.}(2010)\citenamefont
  {Caycedo-Soler}, \citenamefont {Rodr{\'{\i}}guez}, \citenamefont {Quiroga},\
  and\ \citenamefont {Johnson}}]{Caycedo_Soler_2010NJP}%
  \BibitemOpen
  \bibfield  {author} {\bibinfo {author} {\bibfnamefont {F.}~\bibnamefont
  {Caycedo-Soler}}, \bibinfo {author} {\bibfnamefont {F.~J.}\ \bibnamefont
  {Rodr{\'{\i}}guez}}, \bibinfo {author} {\bibfnamefont {L.}~\bibnamefont
  {Quiroga}}, \ and\ \bibinfo {author} {\bibfnamefont {N.~F.}\ \bibnamefont
  {Johnson}},\ }\href {\doibase 10.1088/1367-2630/12/9/095008} {\bibfield
  {journal} {\bibinfo  {journal} {New Journal of Physics}\ }\textbf {\bibinfo
  {volume} {12}},\ \bibinfo {pages} {095008} (\bibinfo {year}
  {2010})}\BibitemShut {NoStop}%
\bibitem [{\citenamefont {Sener}\ \emph {et~al.}(2016)\citenamefont {Sener},
  \citenamefont {Strumpfer}, \citenamefont {Singharoy},\ and\ \citenamefont
  {Hunter}}]{Sener}%
  \BibitemOpen
  \bibfield  {author} {\bibinfo {author} {\bibfnamefont {M.}~\bibnamefont
  {Sener}}, \bibinfo {author} {\bibfnamefont {J.}~\bibnamefont {Strumpfer}},
  \bibinfo {author} {\bibfnamefont {A.}~\bibnamefont {Singharoy}}, \ and\
  \bibinfo {author} {\bibfnamefont {K.}~\bibnamefont {Hunter}, \bibfnamefont
  {C~Neil~Schulten}},\ }\href {\doibase 10.7554/eLife.09541} {\bibfield
  {journal} {\bibinfo  {journal} {eLife}\ }\textbf {\bibinfo {volume} {5}},\
  \bibinfo {pages} {1} (\bibinfo {year} {2016})}\BibitemShut {NoStop}%
\bibitem [{\citenamefont {{Caycedo-Soler}}\ \emph {et~al.}(2010)\citenamefont
  {{Caycedo-Soler}}, \citenamefont {{Rodr\'{\i}guez}}, \citenamefont
  {{Quiroga}},\ and\ \citenamefont {{Johnson}}}]{Caycedo_2010PRL}%
  \BibitemOpen
  \bibfield  {author} {\bibinfo {author} {\bibfnamefont {F.}~\bibnamefont
  {{Caycedo-Soler}}}, \bibinfo {author} {\bibfnamefont {F.~J.}\ \bibnamefont
  {{Rodr\'{\i}guez}}}, \bibinfo {author} {\bibfnamefont {L.}~\bibnamefont
  {{Quiroga}}}, \ and\ \bibinfo {author} {\bibfnamefont {N.~F.}\ \bibnamefont
  {{Johnson}}},\ }\href
  {https://journals.aps.org/prl/abstract/10.1103/PhysRevLett.104.158302}
  {\bibfield  {journal} {\bibinfo  {journal} {Phys. Rev. Lett.}\ }\textbf
  {\bibinfo {volume} {104}},\ \bibinfo {pages} {15832} (\bibinfo {year}
  {2010})}\BibitemShut {NoStop}%
\bibitem [{\citenamefont {Manrique}\ \emph {et~al.}(2016)\citenamefont
  {Manrique}, \citenamefont {Caycedo-Soler}, \citenamefont {De~Mendoza},
  \citenamefont {Rodr\'iguez}, \citenamefont {Quiroga},\ and\ \citenamefont
  {Johnson}}]{ADM1}%
  \BibitemOpen
  \bibfield  {author} {\bibinfo {author} {\bibfnamefont {P.~D.}\ \bibnamefont
  {Manrique}}, \bibinfo {author} {\bibfnamefont {F.}~\bibnamefont
  {Caycedo-Soler}}, \bibinfo {author} {\bibfnamefont {A.}~\bibnamefont
  {De~Mendoza}}, \bibinfo {author} {\bibfnamefont {F.}~\bibnamefont
  {Rodr\'iguez}}, \bibinfo {author} {\bibfnamefont {L.}~\bibnamefont
  {Quiroga}}, \ and\ \bibinfo {author} {\bibfnamefont {N.~F.}\ \bibnamefont
  {Johnson}},\ }\href
  {https://www.sciencedirect.com/science/article/pii/S2211379716304090}
  {\bibfield  {journal} {\bibinfo  {journal} {Results in Physics}\ }\textbf
  {\bibinfo {volume} {6}},\ \bibinfo {pages} {957} (\bibinfo {year}
  {2016})}\BibitemShut {NoStop}%
\bibitem [{\citenamefont {Hu}\ \emph {et~al.}(2002)\citenamefont {Hu},
  \citenamefont {Ritz}, \citenamefont {Damjanovi\'c}, \citenamefont
  {Autenrieth},\ and\ \citenamefont {Schulten}}]{review}%
  \BibitemOpen
  \bibfield  {author} {\bibinfo {author} {\bibfnamefont {X.}~\bibnamefont
  {Hu}}, \bibinfo {author} {\bibfnamefont {T.}~\bibnamefont {Ritz}}, \bibinfo
  {author} {\bibfnamefont {A.}~\bibnamefont {Damjanovi\'c}}, \bibinfo {author}
  {\bibfnamefont {F.}~\bibnamefont {Autenrieth}}, \ and\ \bibinfo {author}
  {\bibfnamefont {K.}~\bibnamefont {Schulten}},\ }\href {\doibase
  10.1017/S0033583501003754} {\bibfield  {journal} {\bibinfo  {journal}
  {Quarterly Reviews of Biophysics}\ }\textbf {\bibinfo {volume} {35}},\
  \bibinfo {pages} {1} (\bibinfo {year} {2002})}\BibitemShut {NoStop}%
\bibitem [{\citenamefont {Jungas}\ \emph {et~al.}(1999)\citenamefont {Jungas},
  \citenamefont {Ranck}, \citenamefont {Rigaud}, \citenamefont {Joliot},\ and\
  \citenamefont {Verméglio}}]{Jungas}%
  \BibitemOpen
  \bibfield  {author} {\bibinfo {author} {\bibfnamefont {C.}~\bibnamefont
  {Jungas}}, \bibinfo {author} {\bibfnamefont {J.-L.}\ \bibnamefont {Ranck}},
  \bibinfo {author} {\bibfnamefont {J.-L.}\ \bibnamefont {Rigaud}}, \bibinfo
  {author} {\bibfnamefont {P.}~\bibnamefont {Joliot}}, \ and\ \bibinfo {author}
  {\bibfnamefont {A.}~\bibnamefont {Verméglio}},\ }\href {\doibase
  10.1093/emboj/18.3.534} {\bibfield  {journal} {\bibinfo  {journal} {The EMBO
  Journal}\ }\textbf {\bibinfo {volume} {18}},\ \bibinfo {pages} {534}
  (\bibinfo {year} {1999})}\BibitemShut {NoStop}%
\bibitem [{\citenamefont {Rousseau}(1975)}]{Rosseau_1975}%
  \BibitemOpen
  \bibfield  {author} {\bibinfo {author} {\bibfnamefont {M.}~\bibnamefont
  {Rousseau}},\ }\href {\doibase 10.1088/0305-4470/8/8/012} {\bibfield
  {journal} {\bibinfo  {journal} {Journal of Physics A: Mathematical and
  General}\ }\textbf {\bibinfo {volume} {8}},\ \bibinfo {pages} {1265}
  (\bibinfo {year} {1975})}\BibitemShut {NoStop}%
\bibitem [{\citenamefont {Rousseau}(1977)}]{Rosseau_1977}%
  \BibitemOpen
  \bibfield  {author} {\bibinfo {author} {\bibfnamefont {M.}~\bibnamefont
  {Rousseau}},\ }\href {\doibase 10.1088/0305-4470/10/6/023} {\bibfield
  {journal} {\bibinfo  {journal} {Journal of Physics A: Mathematical and
  General}\ }\textbf {\bibinfo {volume} {10}},\ \bibinfo {pages} {1043}
  (\bibinfo {year} {1977})}\BibitemShut {NoStop}%
\bibitem [{\citenamefont {Kimble}\ and\ \citenamefont {Mandel}(1984)}]{Kimble}%
  \BibitemOpen
  \bibfield  {author} {\bibinfo {author} {\bibfnamefont {H.}~\bibnamefont
  {Kimble}}\ and\ \bibinfo {author} {\bibfnamefont {L.}~\bibnamefont
  {Mandel}},\ }\href
  {https://journals.aps.org/pra/abstract/10.1103/PhysRevA.30.844} {\bibfield
  {journal} {\bibinfo  {journal} {Physical Review A}\ }\textbf {\bibinfo
  {volume} {30}},\ \bibinfo {pages} {844} (\bibinfo {year} {1984})}\BibitemShut
  {NoStop}%
\bibitem [{\citenamefont {De~Mendoza}\ \emph {et~al.}(2017)\citenamefont
  {De~Mendoza}, \citenamefont {Caycedo-Soler}, \citenamefont {Manrique},
  \citenamefont {Quiroga}, \citenamefont {Rodriguez},\ and\ \citenamefont
  {Johnson}}]{ADM-IOP}%
  \BibitemOpen
  \bibfield  {author} {\bibinfo {author} {\bibfnamefont {A.~M.}\ \bibnamefont
  {De~Mendoza}}, \bibinfo {author} {\bibfnamefont {F.}~\bibnamefont
  {Caycedo-Soler}}, \bibinfo {author} {\bibfnamefont {P.}~\bibnamefont
  {Manrique}}, \bibinfo {author} {\bibfnamefont {L.}~\bibnamefont {Quiroga}},
  \bibinfo {author} {\bibfnamefont {F.~J.}\ \bibnamefont {Rodriguez}}, \ and\
  \bibinfo {author} {\bibfnamefont {N.~F.}\ \bibnamefont {Johnson}},\ }\href
  {http://stacks.iop.org/0953-4075/50/i=12/a=124002} {\bibfield  {journal}
  {\bibinfo  {journal} {Journal of Physics B: Atomic, Molecular and Optical
  Physics}\ }\textbf {\bibinfo {volume} {50}},\ \bibinfo {pages} {124002}
  (\bibinfo {year} {2017})}\BibitemShut {NoStop}%
\bibitem [{\citenamefont {Goh}\ and\ \citenamefont
  {Barab\'asi}(2008)}]{Barabasi}%
  \BibitemOpen
  \bibfield  {author} {\bibinfo {author} {\bibfnamefont {K.-I.}\ \bibnamefont
  {Goh}}\ and\ \bibinfo {author} {\bibfnamefont {A.-L.}\ \bibnamefont
  {Barab\'asi}},\ }\href {http://stacks.iop.org/0295-5075/81/i=4/a=48002}
  {\bibfield  {journal} {\bibinfo  {journal} {EPL (Europhysics Letters)}\
  }\textbf {\bibinfo {volume} {81}},\ \bibinfo {pages} {48002} (\bibinfo {year}
  {2008})}\BibitemShut {NoStop}%
\bibitem [{\citenamefont {Rockower}(1989)}]{Rockower}%
  \BibitemOpen
  \bibfield  {author} {\bibinfo {author} {\bibfnamefont {E.~B.}\ \bibnamefont
  {Rockower}},\ }\href {\doibase 10.1119/1.15957} {\bibfield  {journal}
  {\bibinfo  {journal} {American Journal of Physics}\ }\textbf {\bibinfo
  {volume} {57}},\ \bibinfo {pages} {616} (\bibinfo {year} {1989})}\BibitemShut
  {NoStop}%
\bibitem [{\citenamefont {Chan}\ \emph {et~al.}(2018)\citenamefont {Chan},
  \citenamefont {Gamel}, \citenamefont {Fleming},\ and\ \citenamefont
  {Whaley}}]{Whaley}%
  \BibitemOpen
  \bibfield  {author} {\bibinfo {author} {\bibfnamefont {H.}~\bibnamefont
  {Chan}}, \bibinfo {author} {\bibfnamefont {O.}~\bibnamefont {Gamel}},
  \bibinfo {author} {\bibfnamefont {G.}~\bibnamefont {Fleming}}, \ and\
  \bibinfo {author} {\bibfnamefont {K.}~\bibnamefont {Whaley}},\ }\href
  {https://iopscience.iop.org/article/10.1088/1361-6455/aa9c95/meta} {\bibfield
   {journal} {\bibinfo  {journal} {Journal of Physics B: Atomic, Molecular and
  Optical Physics}\ }\textbf {\bibinfo {volume} {51}},\ \bibinfo {pages} {1}
  (\bibinfo {year} {2018})}\BibitemShut {NoStop}%
\bibitem [{\citenamefont {Trinkunas}\ \emph {et~al.}(2001)\citenamefont
  {Trinkunas}, \citenamefont {Herek}, \citenamefont {Pol\'{\i}vka},
  \citenamefont {Sundstr\"om},\ and\ \citenamefont
  {Pullerits}}]{Trinkunas_PRL}%
  \BibitemOpen
  \bibfield  {author} {\bibinfo {author} {\bibfnamefont {G.}~\bibnamefont
  {Trinkunas}}, \bibinfo {author} {\bibfnamefont {J.~L.}\ \bibnamefont
  {Herek}}, \bibinfo {author} {\bibfnamefont {T.}~\bibnamefont {Pol\'{\i}vka}},
  \bibinfo {author} {\bibfnamefont {V.}~\bibnamefont {Sundstr\"om}}, \ and\
  \bibinfo {author} {\bibfnamefont {T.}~\bibnamefont {Pullerits}},\ }\href
  {\doibase 10.1103/PhysRevLett.86.4167} {\bibfield  {journal} {\bibinfo
  {journal} {Phys. Rev. Lett.}\ }\textbf {\bibinfo {volume} {86}},\ \bibinfo
  {pages} {4167} (\bibinfo {year} {2001})}\BibitemShut {NoStop}%
\bibitem [{\citenamefont {Hess}\ \emph {et~al.}(1995)\citenamefont {Hess},
  \citenamefont {Chachisvilis}, \citenamefont {Timpmann}, \citenamefont
  {Jones}, \citenamefont {Fowler}, \citenamefont {Hunter},\ and\ \citenamefont
  {Sundstr{\"o}m}}]{Hess}%
  \BibitemOpen
  \bibfield  {author} {\bibinfo {author} {\bibfnamefont {S.}~\bibnamefont
  {Hess}}, \bibinfo {author} {\bibfnamefont {M.}~\bibnamefont {Chachisvilis}},
  \bibinfo {author} {\bibfnamefont {K.}~\bibnamefont {Timpmann}}, \bibinfo
  {author} {\bibfnamefont {M.~R.}\ \bibnamefont {Jones}}, \bibinfo {author}
  {\bibfnamefont {G.~J.}\ \bibnamefont {Fowler}}, \bibinfo {author}
  {\bibfnamefont {C.~N.}\ \bibnamefont {Hunter}}, \ and\ \bibinfo {author}
  {\bibfnamefont {V.}~\bibnamefont {Sundstr{\"o}m}},\ }\href
  {https://www.ncbi.nlm.nih.gov/pubmed/11607622} {\bibfield  {journal}
  {\bibinfo  {journal} {Proceedings of the National Academy of Sciences of the
  United States of America}\ }\textbf {\bibinfo {volume} {92}},\ \bibinfo
  {pages} {12333–12337} (\bibinfo {year} {1995})}\BibitemShut {NoStop}%
\bibitem [{\citenamefont {Mattioni}\ \emph {et~al.}(2019)\citenamefont
  {Mattioni}, \citenamefont {Caycedo-Soler}, \citenamefont {Huelga},\ and\
  \citenamefont {Plenio}}]{Mattioni}%
  \BibitemOpen
  \bibfield  {author} {\bibinfo {author} {\bibfnamefont {A.}~\bibnamefont
  {Mattioni}}, \bibinfo {author} {\bibfnamefont {F.}~\bibnamefont
  {Caycedo-Soler}}, \bibinfo {author} {\bibfnamefont {S.~F.}\ \bibnamefont
  {Huelga}}, \ and\ \bibinfo {author} {\bibfnamefont {M.~B.}\ \bibnamefont
  {Plenio}},\ }\href {https://arxiv.org/abs/1812.07905v2.pdf} {\bibfield
  {journal} {\bibinfo  {journal} {arxiv:1812.07905v2
  [physics.cond-mat.mes-hall]}\ } (\bibinfo {year} {2019})}\BibitemShut
  {NoStop}%
\bibitem [{\citenamefont {Ritz}\ \emph {et~al.}(2001)\citenamefont {Ritz},
  \citenamefont {Park},\ and\ \citenamefont {Schulten}}]{ritz}%
  \BibitemOpen
  \bibfield  {author} {\bibinfo {author} {\bibfnamefont {T.}~\bibnamefont
  {Ritz}}, \bibinfo {author} {\bibfnamefont {S.}~\bibnamefont {Park}}, \ and\
  \bibinfo {author} {\bibfnamefont {K.}~\bibnamefont {Schulten}},\ }\href
  {\doibase 10.1021/jp011032r} {\bibfield  {journal} {\bibinfo  {journal} {The
  Journal of Physical Chemistry B}\ }\textbf {\bibinfo {volume} {105}},\
  \bibinfo {pages} {8259} (\bibinfo {year} {2001})}\BibitemShut {NoStop}%
\bibitem [{\citenamefont {Osv\'ath}\ and\ \citenamefont
  {Mar\'oti}(1997)}]{Osvath}%
  \BibitemOpen
  \bibfield  {author} {\bibinfo {author} {\bibfnamefont {S.}~\bibnamefont
  {Osv\'ath}}\ and\ \bibinfo {author} {\bibfnamefont {P.}~\bibnamefont
  {Mar\'oti}},\ }\href {\doibase 10.1016/S0006-3495(97)78130-X} {\bibfield
  {journal} {\bibinfo  {journal} {Biophysical Journal}\ }\textbf {\bibinfo
  {volume} {73}},\ \bibinfo {pages} {972} (\bibinfo {year} {1997})}\BibitemShut
  {NoStop}%
\bibitem [{\citenamefont {{Campillo}}\ \emph {et~al.}(1976)\citenamefont
  {{Campillo}}, \citenamefont {{Shapiro}}, \citenamefont {{Kollman}},
  \citenamefont {{Winn}},\ and\ \citenamefont {{Hyer}}}]{refannihil}%
  \BibitemOpen
  \bibfield  {author} {\bibinfo {author} {\bibfnamefont {A.}~\bibnamefont
  {{Campillo}}}, \bibinfo {author} {\bibfnamefont {S.~L.}\ \bibnamefont
  {{Shapiro}}}, \bibinfo {author} {\bibfnamefont {V.~H.}\ \bibnamefont
  {{Kollman}}}, \bibinfo {author} {\bibfnamefont {K.~R.}\ \bibnamefont
  {{Winn}}}, \ and\ \bibinfo {author} {\bibfnamefont {R.}~\bibnamefont
  {{Hyer}}},\ }\href {https://dx.doi.org/10.1016%2FS0006-3495(76)85666-4}
  {\bibfield  {journal} {\bibinfo  {journal} {Biophys J.}\ }\textbf {\bibinfo
  {volume} {16}},\ \bibinfo {pages} {93} (\bibinfo {year} {1976})}\BibitemShut
  {NoStop}%
\bibitem [{\citenamefont {Haner}\ and\ \citenamefont {Isenor}(1970)}]{Haner}%
  \BibitemOpen
  \bibfield  {author} {\bibinfo {author} {\bibfnamefont {A.~B.}\ \bibnamefont
  {Haner}}\ and\ \bibinfo {author} {\bibfnamefont {N.~R.}\ \bibnamefont
  {Isenor}},\ }\href {\doibase 10.1119/1.1976448} {\bibfield  {journal}
  {\bibinfo  {journal} {American Journal of Physics}\ }\textbf {\bibinfo
  {volume} {38}},\ \bibinfo {pages} {748} (\bibinfo {year} {1970})}\BibitemShut
  {NoStop}%
\bibitem [{\citenamefont {Valencia}\ \emph {et~al.}(2005)\citenamefont
  {Valencia}, \citenamefont {Scarcelli}, \citenamefont {D'Angelo},\ and\
  \citenamefont {Shih}}]{Valencia}%
  \BibitemOpen
  \bibfield  {author} {\bibinfo {author} {\bibfnamefont {A.}~\bibnamefont
  {Valencia}}, \bibinfo {author} {\bibfnamefont {G.}~\bibnamefont {Scarcelli}},
  \bibinfo {author} {\bibfnamefont {M.}~\bibnamefont {D'Angelo}}, \ and\
  \bibinfo {author} {\bibfnamefont {Y.}~\bibnamefont {Shih}},\ }\href {\doibase
  10.1103/PhysRevLett.94.063601} {\bibfield  {journal} {\bibinfo  {journal}
  {Phys. Rev. Lett.}\ }\textbf {\bibinfo {volume} {94}},\ \bibinfo {pages}
  {063601} (\bibinfo {year} {2005})}\BibitemShut {NoStop}%
\bibitem [{\citenamefont {Mashaal}\ and\ \citenamefont
  {Gordon}(2011)}]{Mashaal1}%
  \BibitemOpen
  \bibfield  {author} {\bibinfo {author} {\bibfnamefont {H.}~\bibnamefont
  {Mashaal}}\ and\ \bibinfo {author} {\bibfnamefont {J.~M.}\ \bibnamefont
  {Gordon}},\ }\href {\doibase 10.1364/ol.36.000900} {\bibfield  {journal}
  {\bibinfo  {journal} {Opt. Lett.}\ }\textbf {\bibinfo {volume} {36}},\
  \bibinfo {pages} {900} (\bibinfo {year} {2011})}\BibitemShut {NoStop}%
\bibitem [{\citenamefont {Corkish}\ \emph {et~al.}(2002)\citenamefont
  {Corkish}, \citenamefont {Green},\ and\ \citenamefont {Puzzer}}]{SAreview}%
  \BibitemOpen
  \bibfield  {author} {\bibinfo {author} {\bibfnamefont {R.}~\bibnamefont
  {Corkish}}, \bibinfo {author} {\bibfnamefont {M.}~\bibnamefont {Green}}, \
  and\ \bibinfo {author} {\bibfnamefont {T.}~\bibnamefont {Puzzer}},\ }\href
  {\doibase 10.1016/s0038-092x(03)00033-1} {\bibfield  {journal} {\bibinfo
  {journal} {Solar Energy}\ }\textbf {\bibinfo {volume} {73}},\ \bibinfo
  {pages} {395} (\bibinfo {year} {2002})}\BibitemShut {NoStop}%
\bibitem [{\citenamefont {Boriskina}\ \emph {et~al.}(2016)\citenamefont
  {Boriskina}, \citenamefont {Green}, \citenamefont {Catchpole}, \citenamefont
  {Yablonovitch}, \citenamefont {Beard}, \citenamefont {Okada}, \citenamefont
  {Lany}, \citenamefont {Gershon}, \citenamefont {Zakutayev}, \citenamefont
  {Tahersima}, \citenamefont {Sorger}, \citenamefont {Naughton}, \citenamefont
  {Kempa}, \citenamefont {Dagenais}, \citenamefont {Yao}, \citenamefont {Xu},
  \citenamefont {Sheng}, \citenamefont {Bronstein}, \citenamefont {Rogers},
  \citenamefont {Alivisatos}, \citenamefont {Nuzzo}, \citenamefont {Gordon},
  \citenamefont {Wu}, \citenamefont {Wisser}, \citenamefont {Salleo},
  \citenamefont {Dionne}, \citenamefont {Bermel}, \citenamefont {Greffet},
  \citenamefont {Celanovic}, \citenamefont {Soljacic}, \citenamefont {Manor},
  \citenamefont {Rotschild}, \citenamefont {Raman}, \citenamefont {Zhu},
  \citenamefont {Fan},\ and\ \citenamefont {Chen}}]{Boriskina}%
  \BibitemOpen
  \bibfield  {author} {\bibinfo {author} {\bibfnamefont {S.~V.}\ \bibnamefont
  {Boriskina}}, \bibinfo {author} {\bibfnamefont {M.~A.}\ \bibnamefont
  {Green}}, \bibinfo {author} {\bibfnamefont {K.}~\bibnamefont {Catchpole}},
  \bibinfo {author} {\bibfnamefont {E.}~\bibnamefont {Yablonovitch}}, \bibinfo
  {author} {\bibfnamefont {M.~C.}\ \bibnamefont {Beard}}, \bibinfo {author}
  {\bibfnamefont {Y.}~\bibnamefont {Okada}}, \bibinfo {author} {\bibfnamefont
  {S.}~\bibnamefont {Lany}}, \bibinfo {author} {\bibfnamefont {T.}~\bibnamefont
  {Gershon}}, \bibinfo {author} {\bibfnamefont {A.}~\bibnamefont {Zakutayev}},
  \bibinfo {author} {\bibfnamefont {M.~H.}\ \bibnamefont {Tahersima}}, \bibinfo
  {author} {\bibfnamefont {V.~J.}\ \bibnamefont {Sorger}}, \bibinfo {author}
  {\bibfnamefont {M.~J.}\ \bibnamefont {Naughton}}, \bibinfo {author}
  {\bibfnamefont {K.}~\bibnamefont {Kempa}}, \bibinfo {author} {\bibfnamefont
  {M.}~\bibnamefont {Dagenais}}, \bibinfo {author} {\bibfnamefont
  {Y.}~\bibnamefont {Yao}}, \bibinfo {author} {\bibfnamefont {L.}~\bibnamefont
  {Xu}}, \bibinfo {author} {\bibfnamefont {X.}~\bibnamefont {Sheng}}, \bibinfo
  {author} {\bibfnamefont {N.~D.}\ \bibnamefont {Bronstein}}, \bibinfo {author}
  {\bibfnamefont {J.~A.}\ \bibnamefont {Rogers}}, \bibinfo {author}
  {\bibfnamefont {A.~P.}\ \bibnamefont {Alivisatos}}, \bibinfo {author}
  {\bibfnamefont {R.~G.}\ \bibnamefont {Nuzzo}}, \bibinfo {author}
  {\bibfnamefont {J.~M.}\ \bibnamefont {Gordon}}, \bibinfo {author}
  {\bibfnamefont {D.~M.}\ \bibnamefont {Wu}}, \bibinfo {author} {\bibfnamefont
  {M.~D.}\ \bibnamefont {Wisser}}, \bibinfo {author} {\bibfnamefont
  {A.}~\bibnamefont {Salleo}}, \bibinfo {author} {\bibfnamefont
  {J.}~\bibnamefont {Dionne}}, \bibinfo {author} {\bibfnamefont
  {P.}~\bibnamefont {Bermel}}, \bibinfo {author} {\bibfnamefont {J.-J.}\
  \bibnamefont {Greffet}}, \bibinfo {author} {\bibfnamefont {I.}~\bibnamefont
  {Celanovic}}, \bibinfo {author} {\bibfnamefont {M.}~\bibnamefont {Soljacic}},
  \bibinfo {author} {\bibfnamefont {A.}~\bibnamefont {Manor}}, \bibinfo
  {author} {\bibfnamefont {C.}~\bibnamefont {Rotschild}}, \bibinfo {author}
  {\bibfnamefont {A.}~\bibnamefont {Raman}}, \bibinfo {author} {\bibfnamefont
  {L.}~\bibnamefont {Zhu}}, \bibinfo {author} {\bibfnamefont {S.}~\bibnamefont
  {Fan}}, \ and\ \bibinfo {author} {\bibfnamefont {G.}~\bibnamefont {Chen}},\
  }\href {\doibase 10.1088/2040-8978/18/7/073004} {\bibfield  {journal}
  {\bibinfo  {journal} {Journal of Optics}\ }\textbf {\bibinfo {volume} {18}},\
  \bibinfo {pages} {073004} (\bibinfo {year} {2016})}\BibitemShut {NoStop}%
\bibitem [{\citenamefont {Vandenbosch}\ and\ \citenamefont
  {Ma}(2012)}]{Vandenbosch}%
  \BibitemOpen
  \bibfield  {author} {\bibinfo {author} {\bibfnamefont {G.~A.}\ \bibnamefont
  {Vandenbosch}}\ and\ \bibinfo {author} {\bibfnamefont {Z.}~\bibnamefont
  {Ma}},\ }\href {\doibase 10.1016/j.nanoen.2012.03.002} {\bibfield  {journal}
  {\bibinfo  {journal} {Nano Energy}\ }\textbf {\bibinfo {volume} {1}},\
  \bibinfo {pages} {494} (\bibinfo {year} {2012})}\BibitemShut {NoStop}%
\bibitem [{\citenamefont {Zhao}\ \emph {et~al.}(2018)\citenamefont {Zhao},
  \citenamefont {Gao}, \citenamefont {Cao},\ and\ \citenamefont {Li}}]{Zhao}%
  \BibitemOpen
  \bibfield  {author} {\bibinfo {author} {\bibfnamefont {H.}~\bibnamefont
  {Zhao}}, \bibinfo {author} {\bibfnamefont {H.}~\bibnamefont {Gao}}, \bibinfo
  {author} {\bibfnamefont {T.}~\bibnamefont {Cao}}, \ and\ \bibinfo {author}
  {\bibfnamefont {B.}~\bibnamefont {Li}},\ }\href {\doibase
  10.1364/OE.26.00A178} {\bibfield  {journal} {\bibinfo  {journal} {Opt.
  Express}\ }\textbf {\bibinfo {volume} {26}},\ \bibinfo {pages} {A178}
  (\bibinfo {year} {2018})}\BibitemShut {NoStop}%
\bibitem [{\citenamefont {Scully}\ and\ \citenamefont
  {Zubairy}(1997)}]{Scully}%
  \BibitemOpen
  \bibfield  {author} {\bibinfo {author} {\bibfnamefont {M.~O.}\ \bibnamefont
  {Scully}}\ and\ \bibinfo {author} {\bibfnamefont {M.~S.}\ \bibnamefont
  {Zubairy}},\ }\href {\doibase 10.1017/CBO9780511813993} {\emph {\bibinfo
  {title} {Quantum Optics}}}\ (\bibinfo  {publisher} {Cambridge University
  Press},\ \bibinfo {year} {1997})\BibitemShut {NoStop}%
\bibitem [{\citenamefont {Mandel}(1961)}]{Mandel-CEP}%
  \BibitemOpen
  \bibfield  {author} {\bibinfo {author} {\bibfnamefont {L.}~\bibnamefont
  {Mandel}},\ }\href {\doibase 10.1364/JOSA.51.001342} {\bibfield  {journal}
  {\bibinfo  {journal} {J. Opt. Soc. Am.}\ }\textbf {\bibinfo {volume} {51}},\
  \bibinfo {pages} {1342} (\bibinfo {year} {1961})}\BibitemShut {NoStop}%
\bibitem [{\citenamefont {Mandel}\ and\ \citenamefont {Wolf}(1995)}]{Mandel}%
  \BibitemOpen
  \bibfield  {author} {\bibinfo {author} {\bibfnamefont {L.}~\bibnamefont
  {Mandel}}\ and\ \bibinfo {author} {\bibfnamefont {E.}~\bibnamefont {Wolf}},\
  }\href {\doibase 10.1017/CBO9781139644105} {\emph {\bibinfo {title} {Optical
  Coherence and Quantum Optics}}}\ (\bibinfo  {publisher} {Cambridge University
  Press},\ \bibinfo {year} {1995})\BibitemShut {NoStop}%
\bibitem [{\citenamefont {Zhou}\ \emph {et~al.}(2010)\citenamefont {Zhou},
  \citenamefont {Simon}, \citenamefont {Liu},\ and\ \citenamefont
  {Shih}}]{Shih}%
  \BibitemOpen
  \bibfield  {author} {\bibinfo {author} {\bibfnamefont {Y.}~\bibnamefont
  {Zhou}}, \bibinfo {author} {\bibfnamefont {J.}~\bibnamefont {Simon}},
  \bibinfo {author} {\bibfnamefont {J.}~\bibnamefont {Liu}}, \ and\ \bibinfo
  {author} {\bibfnamefont {Y.}~\bibnamefont {Shih}},\ }\href {\doibase
  10.1103/PhysRevA.81.043831} {\bibfield  {journal} {\bibinfo  {journal} {Phys.
  Rev. A}\ }\textbf {\bibinfo {volume} {81}},\ \bibinfo {pages} {043831}
  (\bibinfo {year} {2010})}\BibitemShut {NoStop}%
\bibitem [{\citenamefont {Liu}\ and\ \citenamefont {Shih}(2009)}]{Shih1}%
  \BibitemOpen
  \bibfield  {author} {\bibinfo {author} {\bibfnamefont {J.}~\bibnamefont
  {Liu}}\ and\ \bibinfo {author} {\bibfnamefont {Y.}~\bibnamefont {Shih}},\
  }\href {\doibase 10.1103/physreva.79.023819} {\bibfield  {journal} {\bibinfo
  {journal} {Physical Review A}\ }\textbf {\bibinfo {volume} {79}} (\bibinfo
  {year} {2009}),\ 10.1103/physreva.79.023819}\BibitemShut {NoStop}%
\bibitem [{\citenamefont {Chen}\ \emph {et~al.}(2011)\citenamefont {Chen},
  \citenamefont {Peng}, \citenamefont {Karmakar}, \citenamefont {Xie},\ and\
  \citenamefont {Shih}}]{Shih2}%
  \BibitemOpen
  \bibfield  {author} {\bibinfo {author} {\bibfnamefont {H.}~\bibnamefont
  {Chen}}, \bibinfo {author} {\bibfnamefont {T.}~\bibnamefont {Peng}}, \bibinfo
  {author} {\bibfnamefont {S.}~\bibnamefont {Karmakar}}, \bibinfo {author}
  {\bibfnamefont {Z.}~\bibnamefont {Xie}}, \ and\ \bibinfo {author}
  {\bibfnamefont {Y.}~\bibnamefont {Shih}},\ }\href {\doibase
  10.1103/physreva.84.033835} {\bibfield  {journal} {\bibinfo  {journal}
  {Physical Review A}\ }\textbf {\bibinfo {volume} {84}} (\bibinfo {year}
  {2011}),\ 10.1103/physreva.84.033835}\BibitemShut {NoStop}%
\bibitem [{\citenamefont {Zardecki}(1971)}]{Zardecki}%
  \BibitemOpen
  \bibfield  {author} {\bibinfo {author} {\bibfnamefont {A.}~\bibnamefont
  {Zardecki}},\ }\href {\doibase 10.1139/p71-206} {\bibfield  {journal}
  {\bibinfo  {journal} {Can. J. Phys.}\ }\textbf {\bibinfo {volume} {49}},\
  \bibinfo {pages} {1724} (\bibinfo {year} {1971})}\BibitemShut {NoStop}%
\bibitem [{\citenamefont {Bures}\ \emph
  {et~al.}(1972{\natexlab{b}})\citenamefont {Bures}, \citenamefont {Delisle},\
  and\ \citenamefont {Zardecki}}]{Bures72}%
  \BibitemOpen
  \bibfield  {author} {\bibinfo {author} {\bibfnamefont {J.}~\bibnamefont
  {Bures}}, \bibinfo {author} {\bibfnamefont {C.}~\bibnamefont {Delisle}}, \
  and\ \bibinfo {author} {\bibfnamefont {A.}~\bibnamefont {Zardecki}},\ }\href
  {\doibase 10.1139/p72-179} {\bibfield  {journal} {\bibinfo  {journal} {Can.
  J. Phys.}\ }\textbf {\bibinfo {volume} {50}},\ \bibinfo {pages} {1307}
  (\bibinfo {year} {1972}{\natexlab{b}})}\BibitemShut {NoStop}%
\bibitem [{\citenamefont {Van~Kampen}(1992)}]{Van_Kampen}%
  \BibitemOpen
  \bibfield  {author} {\bibinfo {author} {\bibfnamefont {N.~G.}\ \bibnamefont
  {Van~Kampen}},\ }\href@noop {} {\emph {\bibinfo {title} {Stochastic processes
  in physics and chemistry}}},\ Vol.~\bibinfo {volume} {1}\ (\bibinfo
  {publisher} {Elsevier},\ \bibinfo {year} {1992})\BibitemShut {NoStop}%
\bibitem [{\citenamefont {B\'edard}(1967)}]{Bedard}%
  \BibitemOpen
  \bibfield  {author} {\bibinfo {author} {\bibfnamefont {G.}~\bibnamefont
  {B\'edard}},\ }\href {\doibase 10.1103/physrev.161.1304} {\bibfield
  {journal} {\bibinfo  {journal} {Phys. Rev.}\ }\textbf {\bibinfo {volume}
  {161}},\ \bibinfo {pages} {1304} (\bibinfo {year} {1967})}\BibitemShut
  {NoStop}%
\bibitem [{\citenamefont {Damjanovi\'c}\ \emph {et~al.}(2000)\citenamefont
  {Damjanovi\'c}, \citenamefont {Ritz},\ and\ \citenamefont
  {Schulten}}]{Damja}%
  \BibitemOpen
  \bibfield  {author} {\bibinfo {author} {\bibfnamefont {A.}~\bibnamefont
  {Damjanovi\'c}}, \bibinfo {author} {\bibfnamefont {T.}~\bibnamefont {Ritz}},
  \ and\ \bibinfo {author} {\bibfnamefont {K.}~\bibnamefont {Schulten}},\
  }\href
  {https://doi.org/10.1002/(SICI)1097-461X(2000)77:1<139::AID-QUA13>3.0.CO;2-S}
  {\bibfield  {journal} {\bibinfo  {journal} {International Journal of Quantum
  Chemistry}\ }\textbf {\bibinfo {volume} {77}},\ \bibinfo {pages} {139}
  (\bibinfo {year} {2000})}\BibitemShut {NoStop}%
\bibitem [{\citenamefont {Monshouwer}\ \emph {et~al.}(1997)\citenamefont
  {Monshouwer}, \citenamefont {Abrahamsson}, \citenamefont {van Mourik},\ and\
  \citenamefont {van Grondelle}}]{vanGrondelle}%
  \BibitemOpen
  \bibfield  {author} {\bibinfo {author} {\bibfnamefont {R.}~\bibnamefont
  {Monshouwer}}, \bibinfo {author} {\bibfnamefont {M.}~\bibnamefont
  {Abrahamsson}}, \bibinfo {author} {\bibfnamefont {F.}~\bibnamefont {van
  Mourik}}, \ and\ \bibinfo {author} {\bibfnamefont {R.}~\bibnamefont {van
  Grondelle}},\ }\href {\doibase 10.1021/jp963377t} {\bibfield  {journal}
  {\bibinfo  {journal} {The Journal of Physical Chemistry B}\ }\textbf
  {\bibinfo {volume} {101}},\ \bibinfo {pages} {7241} (\bibinfo {year}
  {1997})}\BibitemShut {NoStop}%
\bibitem [{\citenamefont {Schroeder}\ \emph {et~al.}(2015)\citenamefont
  {Schroeder}, \citenamefont {Caycedo-Soler}, \citenamefont {Huelga},\ and\
  \citenamefont {Plenio}}]{Schroeder}%
  \BibitemOpen
  \bibfield  {author} {\bibinfo {author} {\bibfnamefont {C.~A.}\ \bibnamefont
  {Schroeder}}, \bibinfo {author} {\bibfnamefont {F.}~\bibnamefont
  {Caycedo-Soler}}, \bibinfo {author} {\bibfnamefont {S.~F.}\ \bibnamefont
  {Huelga}}, \ and\ \bibinfo {author} {\bibfnamefont {M.~B.}\ \bibnamefont
  {Plenio}},\ }\href {\doibase 10.1021/acs.jpca.5b04804} {\bibfield  {journal}
  {\bibinfo  {journal} {The Journal of Physical Chemistry A}\ }\textbf
  {\bibinfo {volume} {119}},\ \bibinfo {pages} {9043} (\bibinfo {year}
  {2015})},\ \bibinfo {note} {pMID: 26256512}\BibitemShut {NoStop}%
\end{thebibliography}

\begin{thebibliography}{18}%
\makeatletter
\providecommand \@ifxundefined [1]{%
 \@ifx{#1\undefined}
}%
\providecommand \@ifnum [1]{%
 \ifnum #1\expandafter \@firstoftwo
 \else \expandafter \@secondoftwo
 \fi
}%
\providecommand \@ifx [1]{%
 \ifx #1\expandafter \@firstoftwo
 \else \expandafter \@secondoftwo
 \fi
}%
\providecommand \natexlab [1]{#1}%
\providecommand \enquote  [1]{``#1''}%
\providecommand \bibnamefont  [1]{#1}%
\providecommand \bibfnamefont [1]{#1}%
\providecommand \citenamefont [1]{#1}%
\providecommand \href@noop [0]{\@secondoftwo}%
\providecommand \href [0]{\begingroup \@sanitize@url \@href}%
\providecommand \@href[1]{\@@startlink{#1}\@@href}%
\providecommand \@@href[1]{\endgroup#1\@@endlink}%
\providecommand \@sanitize@url [0]{\catcode `\\12\catcode `\$12\catcode
  `\&12\catcode `\#12\catcode `\^12\catcode `\_12\catcode `\%12\relax}%
\providecommand \@@startlink[1]{}%
\providecommand \@@endlink[0]{}%
\providecommand \url  [0]{\begingroup\@sanitize@url \@url }%
\providecommand \@url [1]{\endgroup\@href {#1}{\urlprefix }}%
\providecommand \urlprefix  [0]{URL }%
\providecommand \Eprint [0]{\href }%
\providecommand \doibase [0]{http://dx.doi.org/}%
\providecommand \selectlanguage [0]{\@gobble}%
\providecommand \bibinfo  [0]{\@secondoftwo}%
\providecommand \bibfield  [0]{\@secondoftwo}%
\providecommand \translation [1]{[#1]}%
\providecommand \BibitemOpen [0]{}%
\providecommand \bibitemStop [0]{}%
\providecommand \bibitemNoStop [0]{.\EOS\space}%
\providecommand \EOS [0]{\spacefactor3000\relax}%
\providecommand \BibitemShut  [1]{\csname bibitem#1\endcsname}%
\let\auto@bib@innerbib\@empty
\bibitem [{\citenamefont {Mandel}\ and\ \citenamefont {Wolf}(1995)}]{Mandel}%
  \BibitemOpen
  \bibfield  {author} {\bibinfo {author} {\bibfnamefont {L.}~\bibnamefont
  {Mandel}}\ and\ \bibinfo {author} {\bibfnamefont {E.}~\bibnamefont {Wolf}},\
  }\href@noop {} {\emph {\bibinfo {title} {Optical coherence and quantum
  optics}}}\ (\bibinfo  {publisher} {Cambridge university press},\ \bibinfo
  {year} {1995})\BibitemShut {NoStop}%
\bibitem [{\citenamefont {Scully}\ and\ \citenamefont
  {Zubairy}(1997)}]{Scully}%
  \BibitemOpen
  \bibfield  {author} {\bibinfo {author} {\bibfnamefont {M.~O.}\ \bibnamefont
  {Scully}}\ and\ \bibinfo {author} {\bibfnamefont {M.~S.}\ \bibnamefont
  {Zubairy}},\ }\href@noop {} {\emph {\bibinfo {title} {Quantum optics}}}\
  (\bibinfo  {publisher} {Cambridge university press},\ \bibinfo {year}
  {1997})\BibitemShut {NoStop}%
\bibitem [{\citenamefont {Liu}\ and\ \citenamefont {Shih}(2009)}]{Shih1}%
  \BibitemOpen
  \bibfield  {author} {\bibinfo {author} {\bibfnamefont {J.}~\bibnamefont
  {Liu}}\ and\ \bibinfo {author} {\bibfnamefont {Y.}~\bibnamefont {Shih}},\
  }\href {\doibase 10.1103/physreva.79.023819} {\bibfield  {journal} {\bibinfo
  {journal} {Physical Review A}\ }\textbf {\bibinfo {volume} {79}} (\bibinfo
  {year} {2009}),\ 10.1103/physreva.79.023819}\BibitemShut {NoStop}%
\bibitem [{\citenamefont {Chen}\ \emph {et~al.}(2011)\citenamefont {Chen},
  \citenamefont {Peng}, \citenamefont {Karmakar}, \citenamefont {Xie},\ and\
  \citenamefont {Shih}}]{Shih2}%
  \BibitemOpen
  \bibfield  {author} {\bibinfo {author} {\bibfnamefont {H.}~\bibnamefont
  {Chen}}, \bibinfo {author} {\bibfnamefont {T.}~\bibnamefont {Peng}}, \bibinfo
  {author} {\bibfnamefont {S.}~\bibnamefont {Karmakar}}, \bibinfo {author}
  {\bibfnamefont {Z.}~\bibnamefont {Xie}}, \ and\ \bibinfo {author}
  {\bibfnamefont {Y.}~\bibnamefont {Shih}},\ }\href {\doibase
  10.1103/physreva.84.033835} {\bibfield  {journal} {\bibinfo  {journal}
  {Physical Review A}\ }\textbf {\bibinfo {volume} {84}} (\bibinfo {year}
  {2011}),\ 10.1103/physreva.84.033835}\BibitemShut {NoStop}%
\bibitem [{\citenamefont {Zhou}\ \emph {et~al.}(2010)\citenamefont {Zhou},
  \citenamefont {Simon}, \citenamefont {Liu},\ and\ \citenamefont
  {Shih}}]{Shih3}%
  \BibitemOpen
  \bibfield  {author} {\bibinfo {author} {\bibfnamefont {Y.}~\bibnamefont
  {Zhou}}, \bibinfo {author} {\bibfnamefont {J.}~\bibnamefont {Simon}},
  \bibinfo {author} {\bibfnamefont {J.}~\bibnamefont {Liu}}, \ and\ \bibinfo
  {author} {\bibfnamefont {Y.}~\bibnamefont {Shih}},\ }\href {\doibase
  10.1103/PhysRevA.81.043831} {\bibfield  {journal} {\bibinfo  {journal} {Phys.
  Rev. A}\ }\textbf {\bibinfo {volume} {81}},\ \bibinfo {pages} {043831}
  (\bibinfo {year} {2010})}\BibitemShut {NoStop}%
\bibitem [{\citenamefont {Born}\ and\ \citenamefont {Wolf}(1980)}]{Born}%
  \BibitemOpen
  \bibfield  {author} {\bibinfo {author} {\bibfnamefont {M.}~\bibnamefont
  {Born}}\ and\ \bibinfo {author} {\bibfnamefont {E.}~\bibnamefont {Wolf}},\
  }\href@noop {} {\emph {\bibinfo {title} {Principles of optics:
  Electromagnetic Theory of Propagation, Interference and Diffraction of Light
  - 6th edition}}}\ (\bibinfo  {publisher} {Pergamon},\ \bibinfo {year}
  {1980})\BibitemShut {NoStop}%
\bibitem [{\citenamefont {Mandel}(1961)}]{Mandel-CEP}%
  \BibitemOpen
  \bibfield  {author} {\bibinfo {author} {\bibfnamefont {L.}~\bibnamefont
  {Mandel}},\ }\href {\doibase 10.1364/JOSA.51.001342} {\bibfield  {journal}
  {\bibinfo  {journal} {J. Opt. Soc. Am.}\ }\textbf {\bibinfo {volume} {51}},\
  \bibinfo {pages} {1342} (\bibinfo {year} {1961})}\BibitemShut {NoStop}%
\bibitem [{\citenamefont {Bures}\ \emph
  {et~al.}(1972{\natexlab{a}})\citenamefont {Bures}, \citenamefont {Delisle},\
  and\ \citenamefont {Zardecki}}]{Bures71}%
  \BibitemOpen
  \bibfield  {author} {\bibinfo {author} {\bibfnamefont {J.}~\bibnamefont
  {Bures}}, \bibinfo {author} {\bibfnamefont {C.}~\bibnamefont {Delisle}}, \
  and\ \bibinfo {author} {\bibfnamefont {A.}~\bibnamefont {Zardecki}},\ }\href
  {\doibase 10.1139/p72-108} {\bibfield  {journal} {\bibinfo  {journal} {Can.
  J. Phys.}\ }\textbf {\bibinfo {volume} {50}},\ \bibinfo {pages} {760}
  (\bibinfo {year} {1972}{\natexlab{a}})}\BibitemShut {NoStop}%
\bibitem [{\citenamefont {Van~Kampen}(1992)}]{Van_Kampen}%
  \BibitemOpen
  \bibfield  {author} {\bibinfo {author} {\bibfnamefont {N.~G.}\ \bibnamefont
  {Van~Kampen}},\ }\href@noop {} {\emph {\bibinfo {title} {Stochastic processes
  in physics and chemistry}}},\ Vol.~\bibinfo {volume} {1}\ (\bibinfo
  {publisher} {Elsevier},\ \bibinfo {year} {1992})\BibitemShut {NoStop}%
\bibitem [{\citenamefont {Bures}\ \emph
  {et~al.}(1972{\natexlab{b}})\citenamefont {Bures}, \citenamefont {Delisle},\
  and\ \citenamefont {Zardecki}}]{Bures72}%
  \BibitemOpen
  \bibfield  {author} {\bibinfo {author} {\bibfnamefont {J.}~\bibnamefont
  {Bures}}, \bibinfo {author} {\bibfnamefont {C.}~\bibnamefont {Delisle}}, \
  and\ \bibinfo {author} {\bibfnamefont {A.}~\bibnamefont {Zardecki}},\ }\href
  {\doibase 10.1139/p72-179} {\bibfield  {journal} {\bibinfo  {journal} {Can.
  J. Phys.}\ }\textbf {\bibinfo {volume} {50}},\ \bibinfo {pages} {1307}
  (\bibinfo {year} {1972}{\natexlab{b}})}\BibitemShut {NoStop}%
\bibitem [{\citenamefont {Cantrell}\ and\ \citenamefont
  {Fields}(1973)}]{Cantrell}%
  \BibitemOpen
  \bibfield  {author} {\bibinfo {author} {\bibfnamefont {C.~D.}\ \bibnamefont
  {Cantrell}}\ and\ \bibinfo {author} {\bibfnamefont {J.~R.}\ \bibnamefont
  {Fields}},\ }\href {\doibase 10.1103/physreva.7.2063} {\bibfield  {journal}
  {\bibinfo  {journal} {Physical Review A}\ }\textbf {\bibinfo {volume} {7}},\
  \bibinfo {pages} {2063} (\bibinfo {year} {1973})}\BibitemShut {NoStop}%
\bibitem [{\citenamefont {Zardecki}(1971)}]{Zardecki}%
  \BibitemOpen
  \bibfield  {author} {\bibinfo {author} {\bibfnamefont {A.}~\bibnamefont
  {Zardecki}},\ }\href {\doibase 10.1139/p71-206} {\bibfield  {journal}
  {\bibinfo  {journal} {Can. J. Phys.}\ }\textbf {\bibinfo {volume} {49}},\
  \bibinfo {pages} {1724} (\bibinfo {year} {1971})}\BibitemShut {NoStop}%
\bibitem [{\citenamefont {Valencia}\ \emph {et~al.}(2005)\citenamefont
  {Valencia}, \citenamefont {Scarcelli}, \citenamefont {D'Angelo},\ and\
  \citenamefont {Shih}}]{Valencia}%
  \BibitemOpen
  \bibfield  {author} {\bibinfo {author} {\bibfnamefont {A.}~\bibnamefont
  {Valencia}}, \bibinfo {author} {\bibfnamefont {G.}~\bibnamefont {Scarcelli}},
  \bibinfo {author} {\bibfnamefont {M.}~\bibnamefont {D'Angelo}}, \ and\
  \bibinfo {author} {\bibfnamefont {Y.}~\bibnamefont {Shih}},\ }\href {\doibase
  10.1103/PhysRevLett.94.063601} {\bibfield  {journal} {\bibinfo  {journal}
  {Phys. Rev. Lett.}\ }\textbf {\bibinfo {volume} {94}},\ \bibinfo {pages}
  {063601} (\bibinfo {year} {2005})}\BibitemShut {NoStop}%
\bibitem [{\citenamefont {Rockower}(1989)}]{Rockower}%
  \BibitemOpen
  \bibfield  {author} {\bibinfo {author} {\bibfnamefont {E.~B.}\ \bibnamefont
  {Rockower}},\ }\href {\doibase 10.1119/1.15957} {\bibfield  {journal}
  {\bibinfo  {journal} {American Journal of Physics}\ }\textbf {\bibinfo
  {volume} {57}},\ \bibinfo {pages} {616} (\bibinfo {year} {1989})}\BibitemShut
  {NoStop}%
\bibitem [{\citenamefont {Graige}\ \emph {et~al.}(1996)\citenamefont {Graige},
  \citenamefont {Paddock}, \citenamefont {Bruce}, \citenamefont {Feher},\ and\
  \citenamefont {Okamura}}]{Graige}%
  \BibitemOpen
  \bibfield  {author} {\bibinfo {author} {\bibfnamefont {M.~S.}\ \bibnamefont
  {Graige}}, \bibinfo {author} {\bibfnamefont {M.~L.}\ \bibnamefont {Paddock}},
  \bibinfo {author} {\bibfnamefont {J.~M.}\ \bibnamefont {Bruce}}, \bibinfo
  {author} {\bibfnamefont {G.}~\bibnamefont {Feher}}, \ and\ \bibinfo {author}
  {\bibfnamefont {M.~Y.}\ \bibnamefont {Okamura}},\ }\href {\doibase
  10.1021/ja960056m} {\bibfield  {journal} {\bibinfo  {journal} {Journal of the
  American Chemical Society}\ }\textbf {\bibinfo {volume} {118}},\ \bibinfo
  {pages} {9005} (\bibinfo {year} {1996})},\ \Eprint
  {http://arxiv.org/abs/http://dx.doi.org/10.1021/ja960056m}
  {http://dx.doi.org/10.1021/ja960056m} \BibitemShut {NoStop}%
\bibitem [{\citenamefont {Diner}\ \emph {et~al.}(1984)\citenamefont {Diner},
  \citenamefont {Schenck},\ and\ \citenamefont {Vitry}}]{Diner}%
  \BibitemOpen
  \bibfield  {author} {\bibinfo {author} {\bibfnamefont {B.~A.}\ \bibnamefont
  {Diner}}, \bibinfo {author} {\bibfnamefont {C.~C.}\ \bibnamefont {Schenck}},
  \ and\ \bibinfo {author} {\bibfnamefont {D.}~\bibnamefont {Vitry}},\ }\href
  {https://doi.org/10.1016/0005-2728(84)90211-1} {\bibfield  {journal}
  {\bibinfo  {journal} {Biochimica et Biophysica Acta}\ }\textbf {\bibinfo
  {volume} {766}},\ \bibinfo {pages} {9} (\bibinfo {year} {1984})}\BibitemShut
  {NoStop}%
\bibitem [{\citenamefont {Milano}\ \emph {et~al.}(2003)\citenamefont {Milano},
  \citenamefont {Agostiano}, \citenamefont {Mavelli},\ and\ \citenamefont
  {Trotta}}]{Milano}%
  \BibitemOpen
  \bibfield  {author} {\bibinfo {author} {\bibfnamefont {F.}~\bibnamefont
  {Milano}}, \bibinfo {author} {\bibfnamefont {A.}~\bibnamefont {Agostiano}},
  \bibinfo {author} {\bibfnamefont {F.}~\bibnamefont {Mavelli}}, \ and\
  \bibinfo {author} {\bibfnamefont {M.}~\bibnamefont {Trotta}},\ }\href
  {\doibase 10.1046/j.1432-1033.2003.03845.x} {\bibfield  {journal} {\bibinfo
  {journal} {European Journal of Biochemistry}\ }\textbf {\bibinfo {volume}
  {270}},\ \bibinfo {pages} {4595} (\bibinfo {year} {2003})}\BibitemShut
  {NoStop}%
\bibitem [{\citenamefont {Osv\'ath}\ and\ \citenamefont
  {Mar\'oti}(1997)}]{Osvath}%
  \BibitemOpen
  \bibfield  {author} {\bibinfo {author} {\bibfnamefont {S.}~\bibnamefont
  {Osv\'ath}}\ and\ \bibinfo {author} {\bibfnamefont {P.}~\bibnamefont
  {Mar\'oti}},\ }\href {\doibase 10.1016/S0006-3495(97)78130-X} {\bibfield
  {journal} {\bibinfo  {journal} {Biophysical Journal}\ }\textbf {\bibinfo
  {volume} {73}},\ \bibinfo {pages} {972} (\bibinfo {year} {1997})}\BibitemShut
  {NoStop}%
\end{thebibliography}
\end{document}